\documentclass[aps,prl,reprint,superscriptaddress,amsmath,amssymb]{revtex4-2}
\usepackage{graphicx}  
\usepackage{dcolumn}   
\usepackage{bm}        
\usepackage{verbatim}
\usepackage{physics}
\usepackage{orcidlink}
\usepackage{color,xcolor}
\usepackage{CJK}
\graphicspath{{./Figs/}}
\newcommand{\SI}{\textit{SI}}
\begin{document}
\begin{CJK*}{UTF8}{}
\title{Multifractal Scaling in Hi-C Maps}

\author{Seong-Gyu Yang \CJKfamily{mj}{(양성규)}\orcidlink{0000-0003-0583-0006}}
\email{seong-gyu.yang@umu.se}
\affiliation{Integrated Science Lab, Department of Physics, Ume{\aa} University, SE-90187 Ume{\aa}, Sweden}
\affiliation{Department of Physics and Institute of Basic Science, Sungkyunkwan University, Suwon 16419, Republic of Korea}
    
\author{Lucas Hedstr\"om\orcidlink{0000-0002-3315-0633}}
\affiliation{Department of Physics, Ume{\aa} University, SE-90187 Ume{\aa}, Sweden}

\author{Jan Smrek\orcidlink{0000-0003-1764-9298}}
\affiliation{Faculty of Physics, University of Vienna, Boltzmanngasse 5, 1090 Vienna, Austria}

\author{Ludvig Lizana\orcidlink{0000-0003-3174-8145}}
\email{ludvig.lizana@umu.se}
\affiliation{Integrated Science Lab, Department of Physics, Ume{\aa} University, SE-90187 Ume{\aa}, Sweden}
    
\date{\today}

\begin{abstract}
The three-dimensional organization of the genome exhibits rich, scale-dependent structure, as revealed by both chromosome contact maps (e.g., Hi-C maps) and chromatin density measured by microscopy.
Recent studies have reported multifractal scaling in these data.
Yet, the origin of this scaling behavior remains unclear: existing efforts describe it through postulated models.
Here, we show that the multifractal structure of Hi-C maps is a direct consequence of the power-law contact probability $P(s)$, which is itself an empirical observable measured from Hi-C maps.
Starting from $P(s)$ with a single exponent $\gamma$, we analytically derive the mass exponent $\tau(q)$, which characterizes how the $q$-th moment of contact density scales with box size $l$ used to coarse-grain the genomic coordinate.
This multifractal behavior reflects the geometric competition between intra- and inter-segment contacts.
We find that the slope of $\tau(q)$ at large $q$ is given by $2 -\gamma$ when $\gamma <1$, and by $1$ when $\gamma \geq 1$.
We further show that this behavior is robust to noise and consistent across diverse organisms, indicating that it is a universal feature of chromatin organization.
We extend our analysis into double-exponent $P(s)$, and show the $l$ dependence in multifractal behavior.
Taken together, these results provide a physical explanation for multifractal scaling and establish a direct link between the multifractality in Hi-C maps and polymer contact statistics, with the large-$q$ slope of $\tau(q)$ mapping onto a known polymer contact exponent.
\end{abstract}

\maketitle
\end{CJK*}

\section*{Introduction}

The three-dimensional organization of chromatin physically constrains gene regulation and nuclear architecture.
This organization can now be probed directly with modern experimental techniques.
Chromosome conformation capture methods, especially Hi-C, enable genome-wide measurement of chromatin contacts~\cite{lieberman2009comprehensive,rao20143d}, while multipoint FISH images provide complementary measurements of chromosome conformations in three-dimensional space~\cite{wang2016spatial,bintu2018super}.
Together, these approaches reveal structural features such as loops, topologically associated domains (TADs)~\cite{takei2021integrated,takaki2025active}, and chromosomal compartments~\cite{bintu2018super,boettiger2016super,rao2017cohesin,schwarzer2017two,kim2017dynamic,hsieh2022enhancer}.

Beyond these structures, these experimental approaches also reveal continuous scaling behavior across genomic distances.
For example, Hi-C experiments show that the contact probability between two loci at genomic distance $s$ follows $P(s) \sim s^{-\gamma}$ with $\gamma \approx 1$ for $s>1$~Mb~\cite{lieberman2009comprehensive}.
This result has made contact scaling a central observable for understanding chromatin folding.

Polymer physics provides a natural framework for interpreting these scaling laws.
Equilibrium models such as the Gaussian chain or self-avoiding walk predict $P(s) \sim s^{-3/2}$ or $s^{-1.76}$, respectively~\cite{de1979scaling,rubinstein2003polymer}, both inconsistent with the observed $\gamma\approx 1$~\cite{mirny2011fractal}.
This value of $\gamma$ instead points to non-equilibrium or topologically constrained conformations, captured by the crumpled (fractal) globule, characterized by hierarchical self-similar folding without entanglement~\cite{grosberg1993crumpled,mirny2011fractal}.
Regardless of the microscopic mechanism, polymer models share a common form: a power-law contact probability $P(s) \sim s^{-\gamma}$, with the exponent $\gamma$ depending on the model.
This convergence has made $P(s)$ a benchmark observable for evaluating models of chromatin folding.

Multifractal analysis approaches the same contact map data from a different direction.
It characterizes the heterogeneous spatial distribution of local scaling behaviors through a spectrum of exponents~\cite{halsey1986fractal,ott2002chaos,salat2017multifractal}, rather than describing how $P(s)$ varies with $s$, where multiple slopes are observed empirically~\cite{lieberman2009comprehensive,sanborn2015chromatin,chan2024activity}.
Originally developed to describe intermittent energy dissipation in turbulent flows~\cite{frisch1980fully,benzi1984multifractal,she1994universal,balyga1995interpretation,paladin1987degrees,paladin1987anomalous,chhabra1989direct,chhabra1989direct2}, it is now widely used to characterize heterogeneous scaling in diverse systems, including chaotic systems~\cite{ott2002chaos}, brain activity~\cite{wink2008monofractal}, financial markets~\cite{calvet2002multifractality,jiang2019multifractal,jung2020fractality}, and complex networks~\cite{murcio2015multifractal,liu2017fractal}.
When applied to Hi-C maps, it revealed bifractal behavior~\cite{pigolotti2020bifractal}. 
Such multifractal signatures also appear in microscopy images of chromatin density~\cite{LEE2025}, indicating that multifractal scaling is not tied to a single technique. 
Multifractal formalism treats Hi-C maps as geometric objects, much like velocity increments or dissipation fields in turbulence, and analyzes them without reference to polymer physics. 
As a result, the connection between multifractal behavior and polymer scaling remains unclear.
For example, an existing attempt reproduces the multifractal scaling using the hierarchical domain model~\cite{pigolotti2020bifractal}, but it does not explain why or how a polymer with power-law-decaying contacts should exhibit such behavior.

Here, we connect these two descriptions, polymer physics and multifractal formalism.
The key input is the contact probability $P(s)$, which is directly measurable from Hi-C maps and observed to decay as a power law across a broad range of genomic distances.
Starting from this empirical observation, we derive the mass exponent $\tau(q)$ analytically, without assuming any specific polymer conformation.
The only polymer input is this contact probability.
We show that multifractal behavior emerges in Hi-C maps because the two-dimensional organization of the contact map distinguishes contacts near and far from the diagonal, encoding different polymer interactions, namely intra- and inter-segment contacts, whose relative weight changes with the moment order $q$.
This geometric competition alone is sufficient to produce multifractal behavior even from a single power law $P(s)$.
We find that the slope of $\tau(q)$ at large $q$ equals $2 - \gamma$ when $\gamma < 1$, and $1$ when $\gamma \geq 1$.
Specifically, this slope of $\tau(q)$, $2-\gamma$, maps onto a polymer contact exponent: the bulk contact exponent $\beta_b$, governing intra-segment contacts.
This correspondence means that $\tau(q)$, which is directly measurable from Hi-C maps, provides a new route to extract polymer contact exponent $\beta_b$ that is encoded in $P(s)$ without requiring structural modeling of the polymer.

Moreover, our $\tau(q)$ analytically explains the previously reported bifractal behavior~\cite{pigolotti2020bifractal}.
The two asymptotic linear regimes observed in empirical Hi-C maps emerge naturally from the competition between intra- and inter-segment contacts.
Crucially, however, these two regimes are not truly distinct.
They are necessarily connected by a smooth crossover, revealing that the underlying structure is multifractal rather than bifractal.
We test these predictions using Hi-C data from human and mouse cells~\cite{rao20143d} and extend the analysis across multiple species, finding consistent multifractal behavior across diverse organisms.
More broadly, our results demonstrate that $\tau(q)$ carries direct geometric information about the underlying polymer, opening a new window into chromatin organization beyond traditional contact scaling analysis.

%
\section*{Results}
\subsection*{Hi-C Maps from Human and Mouse show Multifractal Behavior}
\begin{figure}[t]
\centering
\includegraphics[width=\linewidth]{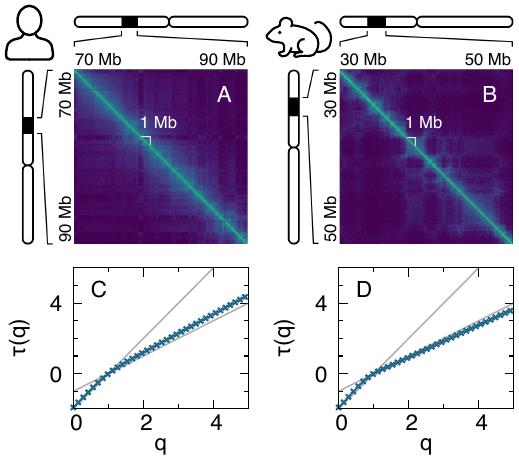}
\caption{
Multifractal analysis results for human (GM12878) and mouse (CH12.LX) chromosome (Chr)~1 Hi-C maps at $25$~kb resolution.
(A and B) Representative parts of the Hi-C maps for (A) human and (B) mouse Chr~1, respectively.
The contact frequencies are KR-normalized and displayed on a logarithmic scale.
The white bracket denotes a $1$~Mb scale.
(C and D) The mass exponent $\tau(q)$ for (C) human and (D) mouse Chr~1 Hi-C maps, respectively, calculated at a box size $l\approx 4$~Mb.
The slope of $\tau(q)$ varies smoothly as $q$ increases, indicating multifractal behavior.
The gray lines show theoretical predictions in the large- and small-$q$ limits, $\tau(q) = q-1$ and $\tau(q) = 2(q-1)$, respectively.
}\label{fig:Empirical}
\end{figure}

We analyze the multifractal behavior of two empirical Hi-C maps ($25$~kb resolution).
One dataset is from human B-lymphoblastoid cells (GM12878), and the other is from mouse B-lymphoblasts (CH12.LX)~\cite{rao20143d}.
After data preprocessing (see \textit{Materials and Methods}), we perform a multifractal analysis by covering the Hi-C maps with boxes of size $l$ and calculating the probability mass $\mu_l$ for each box.
We refer to $\mu_l$ as ``mass'' throughout this paper.
To obtain the generalized partition function $Z(q,l)$, we raise $\mu_l$ to the power $q$ and sum over all boxes, i.e., $Z(q,l) = \sum_\text{box} \mu^q_l$.
As a function of the box size $l$, the generalized partition function scales as $Z(q,l) \sim l^{\tau(q)}$, where the mass exponent $\tau(q)$ captures how different moments of the distribution of $\mu_l$ change with $l$.
A nonlinear $\tau(q)$ is a hallmark of multifractal behavior, often read as distinct scaling in dense versus sparse regions: for Hi-C, we show below that this nonlinearity instead follows from $P(s)$ and the map geometry alone.
Details of the multifractal formalism are provided in \textit{Materials and Methods}.
In \SI, Fig.~S1 depicts the singularity spectrum $f(\alpha)$ for both human and mouse Chr~1 Hi-C maps; the smooth shape of $f(\alpha)$ in both organisms further indicates multifractal rather than bifractal.

Figure~\ref{fig:Empirical} shows $20$~Mb subregions of the Hi-C maps for human and mouse chromosome (Chr)~1 and $\tau(q)$. 
Both maps show a pronounced diagonal corresponding to many local contacts and block-like TADs (see Figs.~\ref{fig:Empirical}A and B).
Figures~\ref{fig:Empirical}C and D show $\tau(q)$ as a function of $q$ when $l\approx 4$~Mb.
Both curves have the same general shape: they exhibit a crossover, where the slope shifts from $2$ for small $q$ to nearly $1$ for large $q$.
The gray lines show the two predicted scaling relations and agree well with the measured $\tau(q)$ (see below).
We selected Chr~1 for both organisms because it is the longest.
In \SI, Figs.~S2--S5 show results for the other chromosomes.

The nonlinear $\tau(q)$ indicates that the Hi-C maps have a more heterogeneous mass distribution than a monofractal.
In such case, the generalized dimension $D_q = \tau(q)/(q-1)$ is independent of $q$ and equal to the fractal dimension $d_f$, yielding a linear $\tau(q) = d_f(q-1)$.
The $q$-dependent slope of $\tau(q)$ observed here, in contrast, reflects the coexistence of multiple scaling regimes.
In the following sections, we explain this scaling behavior analytically using a power-law-decaying contact probability $P(s)$, typical of polymers, without imposing any specific structural model.
Note that we present results only for $q \geq 0$, as $\mu_l^q$ diverges in boxes with zero mass for $q < 0$, making the generalized partition function $Z(q,l)$ ill-defined.

\subsection*{Heuristic argument on origin of multifractal scaling in Hi-C maps}

\begin{figure*}[t!]
\centering
\includegraphics[width=\linewidth]{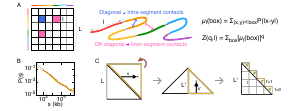}
\caption{
Schematic overview of the multifractal analysis of Hi-C maps.
(A) Relation between box mass $\mu_l(\text{box})$ and polymer contacts.
Diagonal boxes count intra-segment contacts, and off-diagonal boxes count inter-segment contacts.
Here, $l$ is the linear size of the box, which corresponds to the length of each segment.
(B) Contact probability $P(s)$ as a function of genomic distance $s$ for human Chr~1.
$P(s)$ is obtained by applying logarithmic binning to the KR-normalized contact frequencies and exhibits power-law decay with $s$.
The gray dashed line shows a power-law fit with contact exponent $\gamma \approx 1.09$.
(C) Schematic illustration of the matrix transformation used for analytical treatment.
Starting from the upper triangular part of the original contact map of system size $L$, we rotate it by $45$ degrees (red arrow), yielding a right triangle with system size $L' = L/\sqrt{2}$.
Since $P(s)$ of a single-exponent model depends only on the genomic distance $s = |x-y|$, the two halves of the triangle contribute equally to $Z(q,l)$, and we restrict to one half.
We then approximate the hypotenuse of the resulting triangle by a staircase boundary shown as yellow steps and arrows.
The step index $f$ denotes the distance from the diagonal of the original map: $f = 0$ for diagonal boxes and $f > 0$ for off-diagonal boxes.
Each box in the staircase shape has size $l$ and is subdivided into smaller boxes of size $s'_0 = s_0/\sqrt{2}$. 
}\label{fig:scheme}
\end{figure*}

To build intuition, we first present a simple argument, which explains the two slopes in large- and small-$q$ regimes (Figs.~\ref{fig:Empirical}C and D), as well as the crossover, before showing the full derivation.
Detailed calculations are provided in the following section.

In order to capture the two scaling regimes, we decompose $Z(q,l)$ into diagonal and off-diagonal contributions:
\begin{equation} 
	Z(q,l) = Z_\text{diag}(q,l) + Z_\text{off}(q,l),
\end{equation}
where $Z_\text{diag}(q,l)$ accounts for contacts within segments of size $l$, and $Z_\text{off}(q,l)$ contains inter-segment contacts (see Fig.~\ref{fig:scheme}A).
Then, we use the empirically observed form of $P(s)$, which decays as a power law with genomic distance $s$:
\begin{equation}\label{eq:Ps}
P(s) =
\begin{cases}
C,\ &s < s_0,\\
C (s / s_0)^{-\gamma},\ &s \geq s_0,
\end{cases}
\end{equation}
where $C$ is a normalization factor (depends on the system size $L$ but not on $l$), and $s_0$ denotes a cutoff representing the smallest resolvable genomic distance, such as the experimental resolution limit, and prevents divergence at small $s$.
In our empirical data analysis, $s_0 = 25$~kb.
Throughout our analytic calculations, we set $s_0$ to unity by choosing it as the unit of genomic distance.
Figure~\ref{fig:scheme}B shows the empirical contact probability $P(s)$ for human Chr~1.

The slopes in $\tau(q)$ result from a competition between diagonal (intra-segment) and off-diagonal (inter-segment) contributions that dominate $Z(q,l)\sim l^{\tau(q)}$ in different $q$ regimes.
To see this, we note that $\mu_l$ is larger in diagonal boxes than in off-diagonal ones since $\mu_l\sim l \int P(s)ds$ and $P(s)$ decays with $s$. 
Consequently, when evaluating $\mu_l^q$, the diagonal contribution $Z_\text{diag}(q,l)$ dominates for large $q$, while the off-diagonal contribution $Z_\text{off}(q,l)$ does so for small $q$.
For analytical convenience, we approximate the sums defining the mass $\mu_l$ and the generalized partition function $Z(q,l)$ by integrals in the regime $s_0 \ll l \ll L$.

For the diagonal contribution, the integral yields
\begin{equation}\label{eq:contribution_diag}
Z_\text{diag}(q,l) \propto \frac{L}{l} \left( l\int^l_{s_0} s^{-\gamma} ds \right)^q \sim 
\begin{cases}
l^{q-1},\ &\gamma \geq 1,\\
l^{q(2-\gamma)-1},\ &\gamma<1,
\end{cases}
\end{equation}
where the prefactor $L/l$ corresponds to the number of boxes along the diagonal.
For $\gamma \geq 1$, the integral $\int_{s_0}^l s^{-\gamma} ds$ is dominated by the lower cutoff $s_0$ and becomes nearly independent of $l$, leading to $Z_\text{diag}(q,l)\sim l^{q-1}$.
On the other hand, for $\gamma < 1$, the cutoff term becomes negligible, thus $Z_\text{diag}(q,l) \sim l^{q(2-\gamma)-1}$.

As we mentioned above, $Z_\text{diag}(q,l)$ dominates $Z(q,l)$ at large $q$, and thus determines $\tau(q)$.
When $\gamma \geq 1$, we identify $\tau(q)=q-1$.
This is consistent with the empirical Hi-C data, where $\gamma \approx 1.09$ for human Chr~1 and $\gamma \approx 1.1$ for mouse Chr~1 at length scales roughly around $1$~Mb--$10$~Mb.
When $\gamma < 1$, we find $\tau(q) = q(2-\gamma)-1$, i.e., $\tau(q)$'s slope is $2-\gamma$.

The off-diagonal contribution $Z_\text{off}(q,l)$ shows a different scaling behavior than $Z_\text{diag}(q,l)$:
\begin{equation}\label{eq:contribution_off}
Z_\text{off}(q,l) \propto \left(\frac{L}{l}\right)^2\left(d^{-\gamma} l^2 \right)^q \sim l^{2(q-1)},
\end{equation}
where the prefactor $\left( L/l \right)^2$ represents the number of off-diagonal boxes, and $d$ denotes the distance of an off-diagonal box from the diagonal.
Here, we assume that the boxes are sufficiently far away from the diagonal so that the variation of $s$ across them is negligible. 
Under this assumption, the integrated contact probability within each box is proportional to $d^{-\gamma} l^2$.
Finally, since $Z_\text{off}(q,l)$ dominates the small-$q$ regime because of the power-law decaying $P(s)$, we find $\tau(q)=2(q-1)$.

The heuristic argument above considers only the two limiting cases: boxes on the diagonal and those far away from the diagonal.
In practice, however, off-diagonal boxes at intermediate distances interpolate between these extremes.
Their contributions become significant around the crossover $q \sim 1/\gamma$ (derived by equating $\tau(q)$ for large and small $q$) leading to a smooth crossover in $\tau(q)$ rather than a sharp transition.
This contribution persists at large system size $L$, because the number of intermediate boxes increases as $L$ increases.
This gradual transition reflects the shift in the dominant contribution to $Z(q,l)$ from inter- to intra-segment contacts, as $q$ increases.
As a result, the scaling behavior of $\tau(q)$ is inherently smooth, even though it exhibits clear two asymptotic regimes.
%

\subsection*{Analytic derivation of multifractal scaling in the single-exponent model}

This section provides the detailed derivation of the generalized partition function $Z(q,l)$, complementing the heuristic results above.
We consider an $L\times L$ contact map whose entries depend only on the genomic distance $s=|x-y|$ through $P(s)$, where $(x,y)$ is the position of the entry.
Such a map is translationally invariant along the diagonal and carries no position-specific features such as TADs or sub-TADs; $P(s)$ is its only ingredient.
Any multifractality it produces therefore originates from $P(s)$ and the two-dimensional map geometry alone.
For analytical convenience, we rotate the matrix and extend the domain into a staircase geometry (see Fig.~\ref{fig:scheme}C).
Thus, the mass at the $f$-th step from the diagonal is
\begin{equation}\label{eq:our_mu}
\mu_l(f) = \frac{l}{s'_0} \sum_{s' = l f}^{l(f+1)-1} P(s') \approx \frac{l}{s'_0} \int_{lf}^{l(f+1)} P(s') ds',
\end{equation}
where $f$ ranges from $0$ to $L'/l-1$, and $s'_0=s_0/\sqrt{2}$.
Similarly, we express $Z(q,l)$ as
\begin{equation}\label{eq:our_Z}
\begin{split}
&Z(q, l) = 4 \sum_{f=0}^{L'/l-1} \left( \frac{L'}{l} - f \right) \left[ \mu_l(f) \right]^q\\
&\ \approx 4 \frac{L'}{l} \left[\mu_l(f=0) \right]^q + 4\int_{1}^{L'/l} \left(\frac{L'}{l} - f\right) \left[\mu_l(f) \right]^q df,
\end{split}
\end{equation}
where the first term ($f=0$) corresponds to $Z_\text{diag}(q,l)$ and the second one ($f>0$) to $Z_\text{off}(q,l)$.
For simplicity, we redefine $s$ and $L$ by dropping the primes from $s'=s/\sqrt{2}$ and $L'=L/\sqrt{2}$, since this does not affect the $l$-scaling.

We present the calculation results of Eqs.~\ref{eq:our_mu} and \ref{eq:our_Z} for the diagonal and off-diagonal contributions below.
For the diagonal boxes, we compute the mass $\mu_l(f=0)$ using $P(s)$ from Eq.~\ref{eq:Ps} and obtain
\begin{equation}\label{eq:mu_diag}
\mu_l(f=0) = 
\begin{cases}
\frac{C l}{\gamma-1} \left[ \gamma - \left( l/s_0 \right)^{1-\gamma}  \right],  &\gamma \neq 1, \\
C l \left[ 1 + \ln \left(l/s_0 \right) \right], & \gamma = 1.
\end{cases}
\end{equation}
In the first case of Eq.~\ref{eq:mu_diag}, the mass reduces to $\mu_l(f=0) \approx \frac{\gamma Cl}{\gamma-1}$, which scales as $l$ for $\gamma >1$, while for $\gamma < 1$, $\mu_l(f=0) \approx \frac{Cl}{1-\gamma} \left( l/s_0\right)^{1-\gamma}$, yielding $\sim l^{2-\gamma}$.
Substituting these results into Eq.~\ref{eq:our_Z}, we obtain the diagonal contribution as
\begin{equation}\label{eq:detail_Zdiag}
Z_\text{diag}(q,l) \approx
\begin{cases}
\frac{4 \gamma^q C^q }{(\gamma-1)^q} L\, l^{q-1} \sim l^{q-1},\ &\gamma > 1,\\
4 C^q L \left[1 + \ln(l/s_0) \right]^q l^{q-1}\, \sim l^{q-1},\ &\gamma=1,\\
\frac{4 C^q s_0^{q(\gamma-1)}}{(1-\gamma)^q} L \,  l^{q(2-\gamma)-1} \sim l^{q(2-\gamma)-1},\ &\gamma<1.
\end{cases}
\end{equation}
This result is consistent with the heuristic argument in the previous section (see Eq.~\ref{eq:contribution_diag}).

For the off-diagonal region, we obtain the mass $\mu_l(f>0)$ from Eq.~\ref{eq:our_mu} as 
\begin{equation}\label{eq:mu_off}
\mu_l(f>0) \approx C l \left( l/s_0 \right)^{1-\gamma}f^{-\gamma},
\end{equation}
which scales as $l^{2-\gamma}$.
Here, we use the approximations $(f+1)^{1-\gamma} - f^{1-\gamma} \approx (1-\gamma) f^{-\gamma}$ and $\ln(f+1) - \ln f \approx f^{-1}$.
In \SI, Fig.~S6 shows the validation of these approximations.
Using this result, we obtain the dominant term in the off-diagonal contribution as
\begin{equation}\label{eq:detail_Zoff}
Z_\text{off}(q,l) \approx
\begin{cases}
\frac{4  C^q s_0^{q(\gamma-1)}}{q\gamma-1}  L \, l^{q(2-\gamma)-1} \sim l^{q(2-\gamma)-1},\ &q\gamma > 1,\\
4  C^q s_0^{1-q} L \ln(L/l)\, l^{2(q-1)} \sim l^{2(q-1)},\ &q\gamma = 1,\\
\frac{4 C^q s_0^{q(\gamma-1)}}{1-q\gamma} L^{2-q\gamma} \, l^{2(q-1)} \sim l^{2(q-1)},\ &q\gamma < 1.
\end{cases}
\end{equation}
We omit the subleading term $-4\int_1^{L/l} f \left[ \mu_l(f) \right]^q df$, since it does not affect the dominant scaling.
Without this subleading term, the extension to the staircase gives the same result as the extension to the full square on the off-diagonal at the last step in Fig.~\ref{fig:scheme}C.
This also supports the heuristic result in the previous section (see Eq.~\ref{eq:contribution_off}), but with richer behavior: the detailed calculation for $Z_\text{off}(q,l)$ exhibits an additional scaling regime, $Z_\text{off}(q,l) \sim l^{q(2-\gamma)-1}$ when $q\gamma > 1$, which is absent in the heuristic result.
Further details of the derivation are provided in \SI.

\begin{figure}[t!]
\centering
\includegraphics[width=\linewidth]{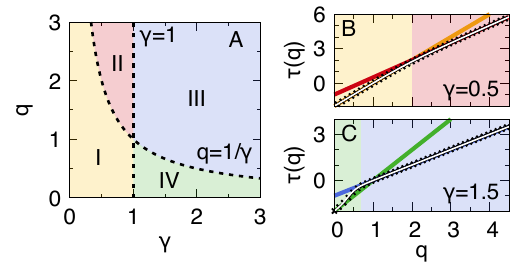}
\caption{
	Region diagram and its validation.
    (A) Diagram of the dominant $\tau(q)$ behavior in $(\gamma,q)$-space.
	The dominant behavior of $\tau(q)$ is divided into four regions:
    (I) yellow: $\tau(q) = 2(q - 1)$;
	(II) red: $\tau(q) = q(2 - \gamma) - 1$;
	(III) blue: $\tau(q) = q - 1$;
    (IV) green: $\tau(q)$ depends on box size $l$:
    $\tau(q) = q-1$ for $l > l^*$, and $\tau(q) = 2(q-1)$ for $l < l^*$, where $l^* \propto (L/s_0)^{(1-q\gamma)/(1-q)}s_0$.
    (B and C) Validation of the analytical results using synthetic contact maps with (B) $\gamma=0.5$ and (C) $\gamma=1.5$.
    The colored lines indicate the analytical predictions from (A), and the white lines show $\tau(q)$ obtained from the analytical expression $Z(q,l) = Z_\text{diag}(q,l) + Z_\text{off}(q,l)$ (see \SI).
    The synthetic maps have system size $L=10^6$ and $\tau(q)$ for the black points in (B) and (C) is calculated with box size $l=10^3$.
}\label{fig:diagram_single}
\end{figure}

By analyzing the dominant contributions to $Z(q,l)$, we find four regimes of $\tau(q)$.
Figure~\ref{fig:diagram_single}A illustrates these regimes in $(\gamma,q)$-space.
The crossover boundaries follow from Eqs.~\ref{eq:detail_Zdiag} and \ref{eq:detail_Zoff}.
When $\gamma < 1$, $\tau(q)$ changes from $\tau(q) = 2(q-1)$ (region I, yellow) to $\tau(q) = q(2-\gamma)-1$ (region II, red) as $q$ increases. 
The crossover is centered at $q = 1/\gamma$, where both contributions become comparable.
In region III (blue), $\tau(q)=q-1$.
In region IV (green), $\tau(q)$ is more complicated since it seems to depend on the box size $l$.
When $l$ is smaller than a certain threshold $l^* \propto (L/s_0)^{(1-q\gamma)/(1-q)}$, $\tau(q)=2(q-1)$ as in region I, while for $l > l^*$, $\tau(q) = q-1$ as in region III.
Since the exponent is positive in region IV, $l^*$ is large, and thus $l < l^*$ typically holds, yielding $\tau(q) = 2(q-1)$ as in region I.

We validate the results using synthetic contact maps of size $L=10^6$ generated from a single power-law contact decay (Eq.~\ref{eq:Ps}), which contain no detailed structure beyond $P(s)$.
Figures~\ref{fig:diagram_single}B and C show the resulting $\tau(q)$ as a function of $q$ for two different values of $\gamma$, where the black dots represent the measurement from the synthetic contact maps.
In Fig.~\ref{fig:diagram_single}B ($\gamma=0.5$), $\tau(q)$ increases with a slope of $2$ until $q = 2$ $(=1/\gamma)$, after which the slope changes to $3/2$ $( = 2-\gamma)$.
For $\gamma=1.5$ (Fig.~\ref{fig:diagram_single}C), the slope of $\tau(q)$ is again $2$ for small $q$ but shifts to $1$ for $q>2/3$.
The colored solid lines indicate the theoretical predictions from Eqs.~\ref{eq:detail_Zdiag} and \ref{eq:detail_Zoff} with colors matching the four regimes in Fig.~\ref{fig:diagram_single}A.
The white lines indicate the results from $Z(q,l) = Z_\text{diag}(q,l) + Z_\text{off}(q,l)$ (see \SI).
We find excellent agreement between theory and simulation.
Overall, $\tau(q)$ in synthetic maps exhibits asymptotically piecewise linear behavior with smooth crossovers between regimes, in agreement with both the heuristic argument and the detailed analytical calculations.


\subsection*{Robustness of Multifractal Behavior to Perturbations}

To assess the robustness of the multifractal behavior against experimental noise, we add white noise $\xi(x,y) \sim \mathcal{N}(0,\sigma^2)$ to each entry $(x,y)$ of empirical Hi-C maps after KR-normalization.
We test two noise levels representing weak and strong perturbations: $\sigma = 10^{-4}$ and $\sigma = 10^{-3}$.
To ensure that the contact maps remain non-negative, we resample $\xi(x,y)$ whenever the resulting entry becomes negative.

\begin{figure}[!t]
\centering
\includegraphics[width=\linewidth]{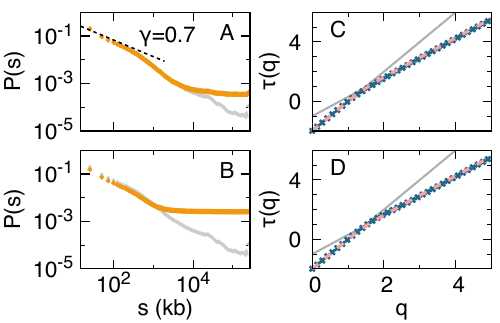}
\caption{
Multifractal analysis of human Chr~1 with additive white noise $\xi(x,y) \sim \mathcal{N}(0,\sigma^2)$ added to each ($x,y$) element of the Hi-C map.
(A and B) The contact probability $P(s)$ exhibits a plateau at large $s$, which is absent in Fig.~\ref{fig:scheme}B.
The contact exponent $\gamma$ in small-$s$ regime is $\gamma\approx 0.7$ extracted from the slope of $\tau(q)$ in the large-$q$ regime $q\in[2,10]$.
(C and D) The mass exponent $\tau(q)$ computed in the box size $l=250$~kb, which lies within the power-law regime of $P(s)$ with exponent $0.7$.
The slope of $\tau(q)$ for large $q$ is approximately $1.3$ $(=2-\gamma)$, and the slope is $2$ for small $q$.
The pink dotted lines show $\tau(q)$ computed from the analytical expression $Z(q,l) = Z_\text{diag}(q,l) + Z_\text{off}(q,l)$ within the single-exponent model.
The noise level $\sigma$ is $\sigma = 10^{-4}$ in (A) and (C), and $\sigma =10^{-3}$ in (B) and (D).
The plateau height in (A and B) is of the same order of magnitude as $\sigma$.
The multifractal scaling remains robust even in the presence of noise.
The black dotted line in (A) shows the slope of $0.7$.
}\label{fig:noise}
\end{figure}

With this additive white noise, $P(s)$ is significantly affected at large $s$ (see Figs.~\ref{fig:noise}A and B).
While $P(s)$ decays with exponent $\gamma\approx 0.7$ in the small-$s$ regime, it exhibits a clear plateau at large $s$.
For comparison, the gray lines show the noise-free $P(s)$, which continues to decay throughout the entire range.
The plateau level increases with $\sigma$, consistent with a signal-independent additive noise that dominates whenever the underlying contact signal becomes smaller than the typical noise amplitude $\sigma$.
The contact exponent $\gamma \approx 0.7$ is extracted from the slope of $\tau(q)$ in the large-$q$ regime.
Using this $\gamma$, we fit the prefactor in $P(s)$ to the empirical $P(s)$ in the small-$s$ regime of $s\in[50~\text{kb}, 250~\text{kb}]$, obtaining a standard error of the fit $2.6\times10^{-5}$.
This small standard error confirms that the contact exponent extracted from $\tau(q)$ accurately describes $P(s)$ in the small-$s$ regime.

In contrast, the mass exponent $\tau(q)$ remains essentially unchanged (Figs.~\ref{fig:noise}C and D, $l=250$~kb).
For both $\sigma$ values, $\tau(q)$ retains its piecewise linear shape, with the same slopes as in the noise-free single-exponent model: the small-$q$ slope is $2$, and the large-$q$ slope is around $1.3 (=2-\gamma)$, consistent with our predictions (see Fig.~\ref{fig:diagram_single} for $\gamma<1$).
The pink dotted lines show the theoretical prediction of $\tau(q)$, which is obtained from the analytical expression of $Z(q,l)$ within the single-exponent model (see \SI).
Note the excellent agreement between the data analysis results (points) and theoretical predictions (lines).
These results imply that the multifractal behavior of the Hi-C maps is robust to additive white noise, as long as the white noise does not erase the power-law decay across the entire $s$ range.
Note that $\tau(q)$ measured here at $l=250$~kb shows a large-$q$ slope of $2-\gamma$ rather than $1$, reflecting that $l$ lies within the small-$s$ regime of $P(s)$ with $\gamma<1$.
This $l$ dependence of $\tau(q)$ and its connection to the two-regime structure of $P(s)$ are analyzed in a later section, where we extend the single-exponent model to a double-exponent model to account for the two-regime structure of $P(s)$.

%


\subsection*{Universal Multifractal Behavior across Diverse Species}
\begin{figure*}[t!]
\centering
\includegraphics[width=\linewidth]{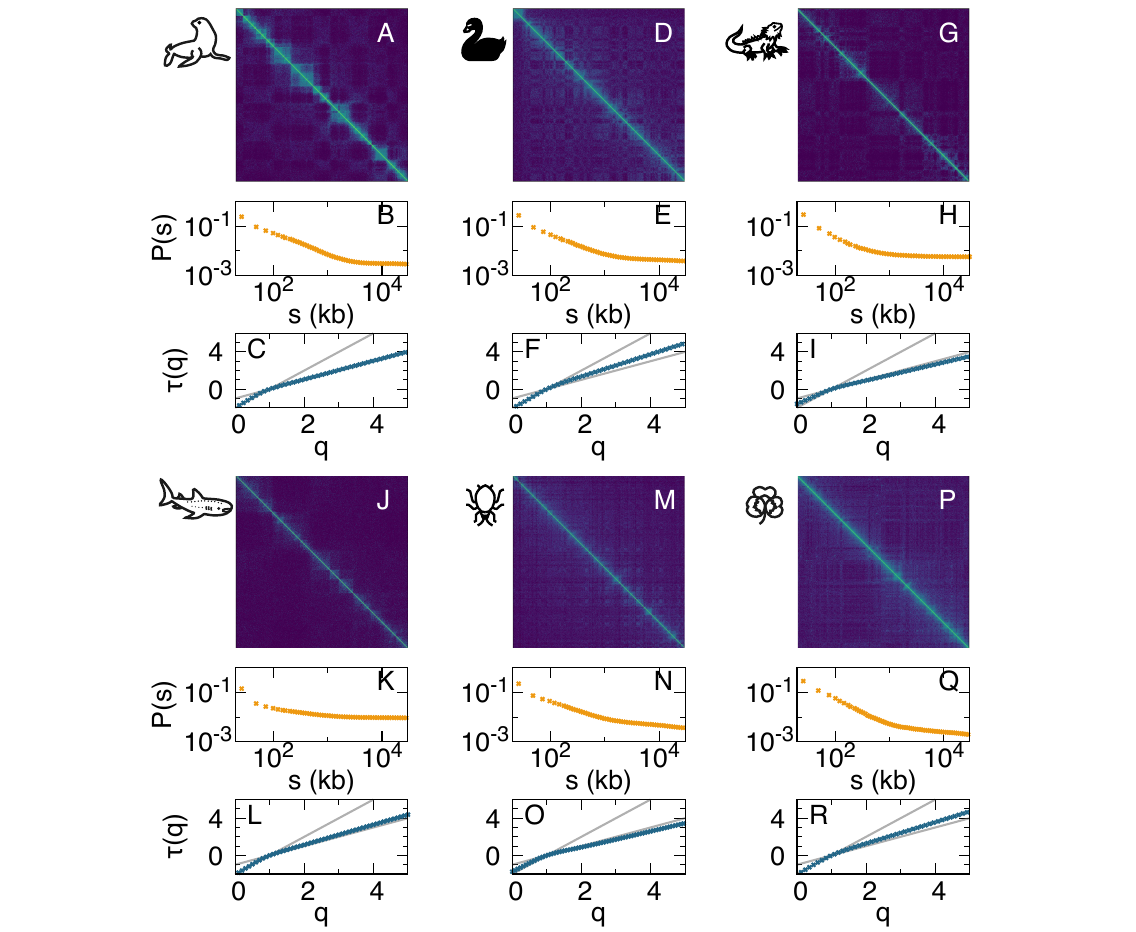}
\caption{   
    Multifractal behavior across six Australian species.
	For each species, we show the Hi-C contact map (top), the contact probability 
$P(s)$ (middle), and the mass exponent $\tau(q)$ (bottom).
	(A--C) New Zealand fur seal, 
	(D--F) black swan, 
	(G--I) central bearded dragon, 
	(J--L) whale shark, 
	(M--O) bluegreen aphid, 
	(P--R) subterranean clover.
	The gray lines in the $\tau(q)$ panels indicate the theoretical predictions $\tau(q) = 2(q-1)$ for small $q$, and $\tau(q)=q-1$ for large $q$.
	All species exhibit a power-law decay in $P(s)$ followed by a large-$s$ plateau, while $\tau(q)$ retains the piece-wise linear form characteristic of multifractal scaling.
	Box size is set to $l\approx 4$~Mb for all species.
    Gray lines show two predictions in the large- and small-$q$ limits.
}\label{fig:others_species}
\end{figure*}

We apply multifractal analysis to six Australian species spanning diverse taxonomic groups: New Zealand fur seal (\textit{Arctocephalus forsteri}, mammal), black swan (\textit{Cygnus atratus}, bird), central bearded dragon (\textit{Pogona vitticeps}, reptile)~\cite{CentralBeardedDragon}, whale shark (\textit{Rhincodon typus}, fish)~\cite{WhaleShark}, bluegreen aphid (\textit{Acyrthosiphon kondoi}, insect), and subterranean clover (\textit{Trifolium subterraneum}, plant)~\cite{Clover}.
The empirical Hi-C maps for these species are publicly available through the Western Australia Genome Atlas~\cite{WAGA}, part of the global DNA Zoo consortium~\cite{dudchenko2017novo,Dudchenko254797}.

For each species, we calculate the contact probability $P(s)$ and the mass exponent $\tau(q)$ from Hi-C data.
The contact probability $P(s)$ exhibits a power-law decay up to intermediate $s$, followed by a plateau for large $s$.
As shown in the previous section, additive noise added to synthetic Hi-C maps produces a similar plateau in $P(s)$ (see Fig.~\ref{fig:noise}).
Regardless of its origin--possibly arising from Rabl-like confinement, finite size effects~\cite{lieberman2009comprehensive,hoencamp20213d,gursoy2014spatial}, or experimental noise--$\tau(q)$ retains the same multifractal scaling we observed in human and mouse Hi-C maps (see Fig.~\ref{fig:Empirical}).
This consistency across diverse taxonomic groups suggests that such multifractal behavior is a universal feature of chromatin architecture.

Figure~\ref{fig:others_species} shows the data for all six species (for the longest Hi-C scaffold): Hi-C maps (top row of each block), contact probabilities $P(s)$ (middle row), and mass exponents $\tau(q)$ (bottom row).
The Hi-C maps exhibit block-diagonal structures, as observed in human and mouse, and $P(s)$ decay as a power law with a plateau at large $s$.
Importantly, $\tau(q)$ exhibits a piecewise-linear shape at intermediate box sizes ($l\approx 4$~Mb).
This confirms that multifractal behavior appears across all six species. 
Including mouse and human suggests that this behavior is universal across organisms.



\subsection*{Double-exponent model explains observed scale dependence of multifractal behavior}

\begin{figure}[t]
\centering
\includegraphics[width=\linewidth]{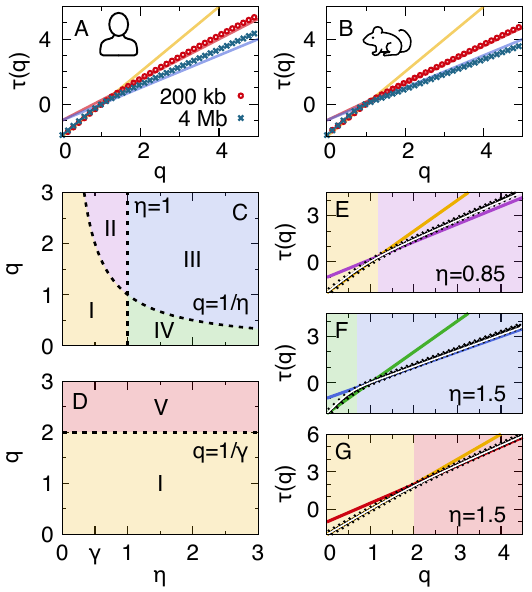}
\caption{
Scale dependence of multifractal behavior and mass exponent $\tau(q)$ in the double-exponent model.
(A and B) The slope of $\tau(q)$ is the same at small $q$ but differs at large $q$ between two box sizes $l=200$~kb and $l=4$~Mb, in both (A) human and (B) mouse Hi-C maps.
Dots represent $\tau(q)$ computed from empirical Hi-C maps, and colored lines show theoretical predictions.
(C and D) Region diagrams show the dominant $\tau(q)$ regions in the (C) large-$l$ ($l>s_T$) and (D) small-$l$ ($l<s_T$) regimes.
(C) For large $l$, the dominant $\tau(q)$ falls into four regions as in the single-exponent model:
(I) yellow: $\tau(q) = 2(q - 1)$;
(II) violet: $\tau(q) = q(2 - \eta) - 1$;
(III) blue: $\tau(q)=q-1$;
(IV) green: $\tau(q)$ depends on the box size $l$: $\tau(q) = q-1$ for $l > l^*$, and $\tau(q) = 2(q-1)$ for $l < l^*$.
The crossover scale is $l^* \propto (L/s_0)^{(1-q\gamma)/(1-q)}s_0$.
(D) For small $l$, the dominant $\tau(q)$ falls into two regions, following the same labeling as in (C):
(I) yellow: $\tau(q) = 2(q - 1)$;
(V) red: $\tau(q) =q(2 - \gamma) - 1$.
(E--G) Validation of the predictions using synthetic contact maps of size $L=10^6$, for (E and F) large $l$ ($l=10^5$) and (G) small $l$ ($l=10^2$).
The transition scale is $s_T=10^3$.
The contact exponent in the large-$s$ regime is $\eta=0.85$ for (E) and $\eta=1.5$ for (F and G), while that in the small-$s$ regime is $\gamma=0.5$.
}\label{fig:diagram_double}
\end{figure}

In the preceding sections, we observed that the slope of $\tau(q)$ in the large-$q$ regime differs between $l\approx 4$~Mb (see Fig.~\ref{fig:Empirical}C) and $l=250$~kb (see Figs.~\ref{fig:noise}B and C), suggesting that this slope depends on the box size $l$.
We examine this dependence more systematically in Figs.~\ref{fig:diagram_double}A and B for human and mouse Hi-C maps, comparing $l=200$~kb (red dots) and $l=4$~Mb (blue dots) along with theoretical predictions: $\tau(q) = 2(q-1)$ (yellow line), $\tau(q) = q(2-\gamma)-1$ (red line), and $\tau(q)=q-1$ (blue line).
The data-derived $\tau(q)$ (red dots) lies close to the prediction $\tau(q)=q(2-\gamma)-1$ (red line) when $l=200$~kb.
However, for $l=4$~Mb (blue dots), it rather follows $\tau(q)=q-1$ (blue line).
This $l$ dependence reflects the two-regime structure of $P(s)$ in empirical Hi-C maps, which the single-exponent model does not capture.

To account for this, we extend our model to a model in which $P(s)$ follows two power laws separated by a transition scale $s_T$.
Such two-regime behavior has been reported in empirical Hi-C maps~\cite{sanborn2015chromatin,rao20143d} and is often associated with loop extrusion and TAD formation~\cite{sanborn2015chromatin,chan2023theory,chan2024activity}.
The contact probability in Fig.~\ref{fig:scheme}B (human) also exhibits a change in its scaling exponent between the short and long genomic distance $s$.

The double-exponent model has two exponents, $\gamma$ and $\eta$, and reads
\begin{equation} \label{eq:p_double}
P(s) =
\begin{cases}
C,&s < s_0,\\
C (s/s_0)^{-\gamma},&s_0 \leq s < s_T,\\
C (s_T/s_0)^{-\gamma}(s/s_T)^{-\eta},&s_T \leq s,
\end{cases}
\end{equation}
where $s_T$ is the transition scale.
The model decays with exponent $\gamma$ in the small-$s$ regime ($s_0 \leq s < s_T$) and with $\eta$ in the large-$s$ regime ($s \geq s_T$).
For simplicity, we approximate the change in the contact exponent as a sharp transition at $s_T$.
Although Hi-C experiments typically yield $\eta \geq 1$, we keep our analysis general and consider both $\eta < 1$ and $\eta \geq 1$, with the assumption $\gamma < 1$.
The small exponent $\gamma<1$ is observed empirically at small genomic distances~\cite{sanborn2015chromatin} (see Figs.~\ref{fig:scheme}B, \ref{fig:noise}A and C for $s \lesssim 500~\text{kb}$).

For this model, $\tau(q)$ depends on the box size $l$ relative to $s_T$.
For large $l$ $(l > s_T)$, each diagonal box spans both the $\gamma$ and $\eta$ regimes of $P(s)$.
Thus, the diagonal contribution $Z_\text{diag}(q,l)$ becomes
\begin{equation}
Z_\text{diag}(q,l) \propto \frac{L}{l} \left[ l\left( \int^{s_T}_{s_0} ds\, s^{-\gamma} +s_T^{\eta-\gamma}\int^{l}_{s_T} ds\, s^{-\eta}\right) \right]^q.
\end{equation}
As in the single-exponent model (with $\eta$ replacing $\gamma$), this term becomes $Z_\text{diag}(q,l) \sim l^{q-1}$ for $\eta \geq 1$, and $Z_\text{diag}(q,l) \sim l^{q(2-\eta)-1}$ for $\eta < 1$.
The off-diagonal contribution $Z_\text{off}(q,l)$ also depends on $\eta$, but its dominant term exhibits the same scaling behavior as in the single-exponent model (see Eq.~\ref{eq:detail_Zoff}) with $\gamma$ replaced by $\eta$, yielding $Z_\text{off}(q,l) \sim l^{q(2-\eta)-1}$ for $q\eta >1$ and $\sim l^{2(q-1)}$ for $q\eta \leq 1$.
Therefore, for large $q$ and large $l$ $(l > s_T)$,  $\tau(q) = q-1$ when $\eta > 1$ and $\tau(q)=q(2-\eta)-1$ when $\eta < 1$.
When $q$ is small, $\tau(q) = 2(q-1)$.
Interestingly, we do not see $\gamma$ dependence in $\tau(q)$ for large $l$ because each diagonal box spans both the $\gamma$ and $\eta$ regimes of $P(s)$, and thus the fine-scale information characterized by $\gamma$ is coarse-grained away by integrating out $P(s)$ over $s$ within $\mu_l$.

Conversely, for small $l$ $(l < s_T)$, each diagonal box lies entirely within the $\gamma$ regime, so $Z_\text{diag}(q,l)$ shows the same scaling as in the single-exponent model, i.e., $Z_\text{diag}(q,l) \sim l^{q(2-\gamma)-1}$ since $\gamma < 1$.
Using the far-from-diagonal approximation, the off-diagonal term $Z_\text{off}(q,l)$ takes the form 
\begin{equation}
Z_\text{off}(q,l) \propto \left(\frac{L}{l}\right)^2 \left( s_T^{\eta-\gamma} d^{-\eta}  l^2 \right)^q \sim l^{2(q-1)},
\end{equation}
where $d$ is the distance from the diagonal.
Thus, for small $l$ $(l< s_T)$, the mass exponent is $\tau(q)=q(2-\gamma)-1$ in the large-$q$ regime, and $\tau(q)=2(q-1)$ in the small-$q$ regime.
The detailed derivations are provided in \SI.
Note that the double-exponent model can also be applied to analyze multifractal behavior of the Hi-C maps of eight species (human and mouse with noise, and six Australian species), whose plateau in large $s$ corresponds to $\eta \approx 0$.

Figure~\ref{fig:diagram_double} shows the resulting diagrams for the scaling regimes associated with the double-exponent model.
For large $l$, the diagram (Fig.~\ref{fig:diagram_double}C) has the same structure as in the single-exponent model (Fig.~\ref{fig:diagram_single}A), with $\eta$ replacing $\gamma$.
Accordingly, the red region in Fig.~\ref{fig:diagram_single}A becomes violet with $\tau(q) = q(2-\eta)-1$.
For small $l$, the diagram (Fig.~\ref{fig:diagram_double}D) contains only two regions: region I (yellow) with $\tau(q)=2(q-1)$ and region V (red) with $\tau(q)=q(2-\gamma)-1$.
We validate these results with synthetic contact maps of size $L=10^6$ by generating data from Eq.~\ref{eq:p_double} (Figs.~\ref{fig:diagram_double}E--G).
The transition scale is set to $s_T=10^3$, and the box size is $l=10^5$ in Figs.~\ref{fig:diagram_double}E and F, and $l=10^2$ in Fig.~\ref{fig:diagram_double}G.
We fix $\gamma=0.5$, and set $\eta=0.85$ for Fig.~\ref{fig:diagram_double}E, and $\eta=1.5$ for Figs.~\ref{fig:diagram_double}F and G.
White lines show $\tau(q)$ calculated from $Z(q,l) = Z_\text{diag}(q,l) + Z_\text{off}(q,l)$ (see \SI).
Again, we note excellent agreement with theoretical predictions.
The small discrepancies between the white lines and colored solid lines in Figs.~\ref{fig:diagram_double}E, F, and G come from the subdominant terms in $\tau(q)$.

\section*{Summary and Discussion}
%
In summary, we have observed multifractal behavior in Hi-C maps from human and mouse cells by measuring the mass exponent $\tau(q)$.
As a minimal description, we have adopted a single-exponent model in which the contact probability $P(s)$ decays as a power law, motivated by the empirical observations that $P(s)$ decays as a power law in Hi-C maps~\cite{lieberman2009comprehensive,nagano2013single,sanborn2015chromatin}, consistent with polymer scaling theory~\cite{halverson2011molecular,mirny2011fractal,grosberg2014annealed}.
Using this model, we have revealed that the multifractality arises from the competition between diagonal and off-diagonal contributions, with the diagonal dominating at large $q$ and the off-diagonal at small $q$.
We have validated our results with synthetic Hi-C maps, which show excellent agreement with our theoretical predictions.
We have also shown that multifractal behavior is robust to noise and observed across several species.
Our findings suggest that multifractal behavior is a universal feature of chromosome contact maps.

For a given polymer model with contact exponent $\gamma$, our results provide a direct prediction of the multifractal scaling of the polymer model, establishing a direct link between multifractal behavior and polymer physics.
The mass $\mu_l$ in the diagonal and off-diagonal boxes corresponds to two distinct types of polymer contacts: intra-segment (diagonal) and inter-segment (off-diagonal).
These are governed by two contact-associated exponents in polymer physics, which we refer to as the bulk contact exponent $\beta_b$ and the surface contact exponent $\beta_s$.
While $\beta_b$ characterizes the scaling of the number of contacts within a segment [$\beta_b=\gamma_c$ in Ref.~\cite{rosa2017beyond}], $\beta_s$ characterizes the scaling of the number of contacts between two different segments [$\beta_s=\beta$ in Refs.~\cite{halverson2011molecular,smrek2013novel,halverson2014melt,rosa2017beyond}].
Each contribution to $Z(q,l)$ encodes one of these exponents: $Z_\text{diag}(q,l)$ encodes $\beta_b$ through the scaling of intra-segment contacts, while $Z_\text{off}(q,l)$ encodes $\beta_s$ through the scaling of inter-segment contacts.

However, in the full partition function $Z(q,l) = Z_\text{diag}(q,l) + Z_\text{off}(q,l)$, only the dominant contribution determines the leading $\tau(q)$, and these two exponents are not equally accessible.
The slope of $\tau(q)$ in the large-$q$ regime directly encodes $\beta_b$: it equals $2-\gamma$ when $\gamma<1$, and $1$ when $\gamma \geq 1$, consistent with the dominance of $Z_\text{diag}(q,l)$ at large $q$.
In contrast, $\beta_s$, which equals $1$ when $\gamma \leq 1$, and $2-\gamma$ when $\gamma > 1$, is hidden from the slope of $\tau(q)$ through two mechanisms.
First, $\beta_s$ appears in $Z_\text{off}(q,l)$ only when $q\gamma > 1$ (i.e., $q > 1/\gamma$), but this is precisely the regime where $Z_\text{diag}(q,l)$ already dominates (region II in Fig.~\ref{fig:diagram_single}), masking $\beta_s$ by the stronger intra-segment contribution.
Second, when $q\gamma < 1$, $Z_\text{off}(q,l)$ dominates, but the slope of $\tau(q)$ becomes $2$, independent of $\gamma$, because the off-diagonal boxes scale as $l^{-2}$ in number and each contains $O(l^2)$ entries, reflecting the two-dimensional geometry of the Hi-C contact map.
Through these two mechanisms, $\beta_s$ is hidden from the slope of $\tau(q)$ across the entire range of $q$.
As a result, the multifractal behavior of Hi-C maps selectively encodes $\beta_b$, while $\beta_s$, though present in the underlying contact statistics through $Z_\text{off}(q,l)$, is not directly readable from the slope of $\tau(q)$.
How to extract $\beta_s$ directly from Hi-C multifractal analysis thus remains an open question.

A contact exponent $\gamma < 1$ has been observed in double-folded ring polymers, with $\gamma \approx 0.75$~\cite{rosa2019conformational}.
This small contact exponent $\gamma$ is also observed within TADs~\cite{sanborn2015chromatin,fudenberg2016formation}, where loop extrusion generates locally compact conformations.
The $\gamma <1$ regime in the double-exponent model thus directly corresponds to the small-scale organization of chromatin shaped by loop extrusion.

While the exponents $\gamma$, $\beta_s$ and $\beta_b$ characterize the contacts, spatial conformation of the polymer is described by the exponent $\nu=1/d_{f}$ governing the scaling of the end-to-end distance $R$ of a segment of length $s$ as $R\sim s^{1/d_{f}}$, where $d_f$ is the fractal dimension.
Although the exponent $\gamma$ is constrained but not determined by $d_{f}$ in general~\cite{halverson2014melt}, in a mean-field approximation in 3D space $\gamma = 3/d_{f}$.
From this perspective, $\gamma < 1$ indicates polymer conformation with $d_{f} > 3$~\cite{chan2024activity}, which has been associated with active polymer dynamics~\cite{chan2024activity}.
Note that such scaling can hold only up to some finite scale, since otherwise the density would diverge, which is unphysical.
An effective fractal dimension higher than $3$ (within the mean-field interpretation) further indicates that the segment folds compactly on itself, so that intra-segment contacts dominate.

To the best of our knowledge, the functional form of the mass exponent $\tau(q)$ has historically been obtained from phenomenological arguments rather than being derived from microscopic mechanisms.
In turbulence, the multifractal behavior is captured by cascade models~\cite{benzi1984multifractal,she1994universal}, and in financial market data by multiplicative cascade models~\cite{calvet2002multifractality}.
In those systems, the multifractal behavior encodes an irreducible hierarchy of scaling exponents that is not fully determined by two-point statistics.
However, in Hi-C maps, the situation is fundamentally different.
We have shown that the observed multifractal behavior is fully determined by the power-law contact probability $P(s)$ and the two-dimensional organization of the contact map, without requiring any cascade-like hierarchical structure.
The multifractality of Hi-C maps is therefore reducible to two-point statistics.
We stress that reducible does not mean spurious or apparent in the previous studies~\cite{jiang2019multifractal,kantelhardt2002multifractal}.
Unlike the finite-size artifacts that vanish as the system size grows, this reducible multifractality is real and robust, surviving large system sizes and added noise.
It is simply not an independent signature of a multiplicative cascade.

We emphasize that this classification reflects the nature of Hi-C measurements rather than the absence of hierarchical organization in genome folding itself.
Genomic folding is known to exhibit cascade-like hierarchical structures, ranging from chromosome territories and compartments to TADs, sub-TADs, and chromatin loops~\cite{gibcus2013hierarchy,lieberman2009comprehensive,rao20143d,sanborn2015chromatin,fudenberg2016formation,bernenko2023mapping}.
However, Hi-C maps encode only pairwise contact frequencies between two loci, and such two-point statistics are, by construction, blind to the higher-order correlations.
A genuine cascade structure, irreducible to two-point contacts, would manifest through such higher-order correlations.

%
Recently, multifractal behavior has also been reported in real-space images of cell nuclei~\cite{LEE2025}.
The authors concluded that multifractal behavior reflects the heterogeneous spatial distribution of chromatin density in the nucleus.
The multifractality observed in our study, however, has a fundamentally different origin.
Our analysis is performed on the Hi-C maps, which encode contact statistics between genomic loci as a function of genomic distance, rather than the spatial distribution of chromatin density.
As a result, the multifractal behavior in our study arises from the coexistence of two distinct polymer scaling regimes rather than from spatial heterogeneity.
But both perspectives are connected and provide complementary insights.
While the image-based multifractal analysis captures the heterogeneous packing of chromatin in real space, our contact-based analysis reveals the underlying polymer physics scaling governing chromatin organization.

In addition to advancing our understanding of chromatin folding, our multifractal analysis provides a computationally reliable method for extracting the contact exponents $\gamma$ and $\eta$ from Hi-C data.
Existing approaches obtain these exponents either by fitting $P(s)$ directly, by log-derivative methods, or by fitting the measured $\tau(q)$ to an assumed structural model.
The last route was taken by a previous study~\cite{pigolotti2020bifractal}, which fit $\tau(q)$ (denoted $K(q)$ in their study) to a hierarchical domain model characterized by a single parameter $a$, yielding $\gamma= \ln(4a)/\ln(2)$.
In contrast, our framework reads $\gamma$ directly from the slope of $\tau(q)$ without assumed structural model.
Our method also provides an integral-based estimator that is potentially less sensitive to noise than direct fitting of log-derivative methods.
Moreover, by varying the box size $l$, our framework naturally accommodates the two-regime contact decay observed empirically, in which $P(s)$ crosses over from $\gamma$ to $\eta$.
This scale-resolved analysis suggests that both exponents $\gamma$ and $\eta$ can be extracted from $\tau(q)$ within a single framework.

In conclusion, the multifractal behavior of Hi-C maps arises from the geometric competition between intra- and inter-segment contacts in a polymer under a power-law-decaying contact probability.
We anticipate that this competition is a general feature of eukaryotic chromatin organization, as supported by our cross-species analysis spanning diverse taxonomic groups.


\section*{Materials and Methods}

\subsection*{Data preprocessing}
We preprocess human and mouse Hi-C data (GSE63525)~\cite{rao20143d} by removing centromeric loci, which exhibit zero contact frequency in the corresponding rows and columns of the Hi-C maps.
The contact frequencies are then normalized using KR normalization~\cite{knight2013fast} to eliminate technical biases.
For Hi-C maps of six species from the Western Australia Genome Atlas~\cite{dudchenko2017novo,WAGA}, we use the hic-straw~\cite{DURAND201699} Python package to read the Hi-C data and apply KR normalization to each Hi-C scaffold.
We use all Hi-C maps at $25$~kb resolution throughout this study.

\subsection*{Multifractal formalism}
Multifractal formalism is a tool that characterizes how local singularity exponents are distributed across the system~\cite{halsey1986fractal,jiang2019multifractal,salat2017multifractal,ott2002chaos}.
We apply this formalism to analyze the scaling behavior of Hi-C maps.
First, we partition the Hi-C map into boxes of size $l$.
The probability mass in a box $b_i$ is defined as
\begin{equation}\label{eq:OneBox_Mass}
    \mu_l(b_i) = \sum_{(x,y)\in b_i} P(|x-y|),
\end{equation}
where the sum runs over all pairs of loci $(x,y)$ within box $b_i$, and $P(|x-y|)$ is the normalized contact probability between the pair $(x,y)$.
Note that $P(|x-y|)$ is normalized such that $\sum_{b_i}\sum_{(x,y)\in b_i} P(|x-y|) = 1$.
The mass $\mu_l(b_i)$ quantifies the local concentration of contacts within box $b_i$.
The mass $\mu_l$ in in diagonal boxes represents intra-segment contacts, while that in off-diagonal boxes corresponds to inter-segment contacts.

From the mass $\mu_l$ over all boxes, the generalized partition function is defined as
\begin{equation}\label{eq:Z_general}
    Z(q, l) = \sum_{i} [\mu_l(b_i)]^q,
\end{equation}
which quantifies the $q$-th moment of the mass distribution.
In multifractal formalism, $Z(q,l)$ is assumed to scale as 
\begin{equation}\label{eq:Z_scaling}
    Z(q, l) \sim l^{\tau(q)},
\end{equation}
where $\tau(q)$ is the mass exponent.
A linear $\tau(q)$ indicates monofractal behavior, while a nonlinear $\tau(q)$ represents multifractal behavior.

In the multifractal formalism, the mass in each box is assumed to scale as $\mu_l(b_i) \sim l^{\alpha_i}$, where $\alpha_i$ is a local scaling exponent that varies across boxes.
The exponent $\alpha_i$ is also referred to as the H{\"o}lder exponent or singularity index.
The mass exponent $\tau(q)$ is one of the central quantities characterizing the multifractal structure.
The complementary quantity is the singularity spectrum $f(\alpha)$, which is related to $\tau(q)$ via a Legendre-Fenchel transform and contains equivalent information~\cite{falconer1994multifractal,chhabra1989direct,chhabra1989direct2,halsey1986fractal,salat2017multifractal}.
In this study, we focus on $\tau(q)$ since its behavior directly reveals the competition between diagonal and off-diagonal contributions.
The mass exponent $\tau(q)$ and the singularity spectrum $f(\alpha)$ are shown in \SI: for human and mouse Chr~1 Hi-C maps (\SI, Fig.~S1), 
for the single-exponent model (\SI, Fig.~S7), and for the double-exponent model (\SI, Fig.~S8).
Note that the generalized fractal dimension $D_q$ can be obtained from $\tau(q)$ as $D_q = \tau(q)/(q-1)$, which corresponds to the box-counting dimension at $q=0$, the information dimension at $q=1$, and the correlation dimension at $q=2$.

\subsection*{Numerical calculation}
For a given box size $l$, we first partition the Hi-C map into boxes of size $l\times l$ and compute the mass $\mu_l(b_i)$ for each box $b_i$ as the sum of the normalized contact probabilities within the box.
We then compute the generalized partition function using Eq.~\ref{eq:Z_general}, and the calculation is repeated for various values of $q$ and $l$.

From the resulting $Z(q,l)$, we numerically calculate the mass exponent $\tau(q)$ at a given $l$ as the logarithmic derivative:
\begin{equation}\label{eq:log_derivative}
    \tau(q) = \frac{d}{d \ln l}\ln Z(q,l).
\end{equation}
Within the scaling regime where $Z(q,l)\sim l^{\tau(q)}$, we have $\ln Z(q,l) = \tau(q) \ln l + C(q)$, where $C(q)$ is independent of $l$.
The ratio $\ln Z(q,l)/\ln l$ contains an additional term $C(q)/\ln l$, which vanishes only as $l\to \infty$ for $\forall q$.
In contrast, the derivative eliminates $C(q)$ and yields $\tau(q)$ directly.
To compute the derivative with respect to $\ln l$ numerically, we employ Fornberg's finite difference algorithm~\cite{fornberg1988generation}. Specifically, we use a five-point stencil at each evaluation point on the $\ln l$ grid, which yields fourth-order accuracy.
Fornberg's algorithm allows accurate derivatives even when the grid is non-uniformly spaced.
In \SI, Fig.~S9, we compare the ratio $\ln Z(q,l) / \ln l$ and the mass exponent $\tau(q)$ in Eq.~\ref{eq:log_derivative} by varying $l$.
The difference scales as $1/\ln l$

%
\section*{Acknowledgments}
The authors thank A.~Grosberg and A.~Rosa for fruitful discussions.
This work was supported by Basic Science Research Program through the National Research Foundation of Korea (NRF) funded by the Ministry of Education (Grant No.~RS-2025-02312897) and the Swedish Research Council (Vetenskapsr{\aa}det) (Grants No.~2021-04080 and No.~2022-06543).
Unpublished genome assemblies and sequencing data for six Australian species are used with permission from the DNA Zoo Consortium (\url{dnazoo.org}).

\section*{Data Availability}
Human and mouse Hi-C maps used in this study are publicly available in the Gene Expression Omnibus (GEO) database, \url{www.ncbi.nlm.nih.gov/geo/} (accession No.~GSE63525)~\cite{rao20143d}.
Hi-C maps for the other six species are found in DNA Zoo, Western Australia Genome Atlas~\cite{dudchenko2017novo,WAGA}.
Analysis codes have been deposited in GitHub~\cite{our_git}.
%
%
%

\bibliographystyle{apsrev4-2}

\begin{thebibliography}{61}%
\makeatletter
\providecommand \@ifxundefined [1]{%
 \@ifx{#1\undefined}
}%
\providecommand \@ifnum [1]{%
 \ifnum #1\expandafter \@firstoftwo
 \else \expandafter \@secondoftwo
 \fi
}%
\providecommand \@ifx [1]{%
 \ifx #1\expandafter \@firstoftwo
 \else \expandafter \@secondoftwo
 \fi
}%
\providecommand \natexlab [1]{#1}%
\providecommand \enquote  [1]{``#1''}%
\providecommand \bibnamefont  [1]{#1}%
\providecommand \bibfnamefont [1]{#1}%
\providecommand \citenamefont [1]{#1}%
\providecommand \href@noop [0]{\@secondoftwo}%
\providecommand \href [0]{\begingroup \@sanitize@url \@href}%
\providecommand \@href[1]{\@@startlink{#1}\@@href}%
\providecommand \@@href[1]{\endgroup#1\@@endlink}%
\providecommand \@sanitize@url [0]{\catcode `\\12\catcode `\$12\catcode
  `\&12\catcode `\#12\catcode `\^12\catcode `\_12\catcode `\%12\relax}%
\providecommand \@@startlink[1]{}%
\providecommand \@@endlink[0]{}%
\providecommand \url  [0]{\begingroup\@sanitize@url \@url }%
\providecommand \@url [1]{\endgroup\@href {#1}{\urlprefix }}%
\providecommand \urlprefix  [0]{URL }%
\providecommand \Eprint [0]{\href }%
\providecommand \doibase [0]{https://doi.org/}%
\providecommand \selectlanguage [0]{\@gobble}%
\providecommand \bibinfo  [0]{\@secondoftwo}%
\providecommand \bibfield  [0]{\@secondoftwo}%
\providecommand \translation [1]{[#1]}%
\providecommand \BibitemOpen [0]{}%
\providecommand \bibitemStop [0]{}%
\providecommand \bibitemNoStop [0]{.\EOS\space}%
\providecommand \EOS [0]{\spacefactor3000\relax}%
\providecommand \BibitemShut  [1]{\csname bibitem#1\endcsname}%
\let\auto@bib@innerbib\@empty
\bibitem [{\citenamefont {Lieberman-Aiden}\ \emph {et~al.}(2009)\citenamefont
  {Lieberman-Aiden}, \citenamefont {Van~Berkum}, \citenamefont {Williams},
  \citenamefont {Imakaev}, \citenamefont {Ragoczy}, \citenamefont {Telling},
  \citenamefont {Amit}, \citenamefont {Lajoie}, \citenamefont {Sabo},
  \citenamefont {Dorschner} \emph {et~al.}}]{lieberman2009comprehensive}%
  \BibitemOpen
  \bibfield  {author} {\bibinfo {author} {\bibfnamefont {E.}~\bibnamefont
  {Lieberman-Aiden}}, \bibinfo {author} {\bibfnamefont {N.~L.}\ \bibnamefont
  {Van~Berkum}}, \bibinfo {author} {\bibfnamefont {L.}~\bibnamefont
  {Williams}}, \bibinfo {author} {\bibfnamefont {M.}~\bibnamefont {Imakaev}},
  \bibinfo {author} {\bibfnamefont {T.}~\bibnamefont {Ragoczy}}, \bibinfo
  {author} {\bibfnamefont {A.}~\bibnamefont {Telling}}, \bibinfo {author}
  {\bibfnamefont {I.}~\bibnamefont {Amit}}, \bibinfo {author} {\bibfnamefont
  {B.~R.}\ \bibnamefont {Lajoie}}, \bibinfo {author} {\bibfnamefont {P.~J.}\
  \bibnamefont {Sabo}}, \bibinfo {author} {\bibfnamefont {M.~O.}\ \bibnamefont
  {Dorschner}}, \emph {et~al.},\ }\href@noop {} {\bibfield  {journal} {\bibinfo
   {journal} {Science}\ }\textbf {\bibinfo {volume} {326}},\ \bibinfo {pages}
  {289} (\bibinfo {year} {2009})}\BibitemShut {NoStop}%
\bibitem [{\citenamefont {Rao}\ \emph {et~al.}(2014)\citenamefont {Rao},
  \citenamefont {Huntley}, \citenamefont {Durand}, \citenamefont {Stamenova},
  \citenamefont {Bochkov}, \citenamefont {Robinson}, \citenamefont {Sanborn},
  \citenamefont {Machol}, \citenamefont {Omer}, \citenamefont {Lander} \emph
  {et~al.}}]{rao20143d}%
  \BibitemOpen
  \bibfield  {author} {\bibinfo {author} {\bibfnamefont {S.~S.}\ \bibnamefont
  {Rao}}, \bibinfo {author} {\bibfnamefont {M.~H.}\ \bibnamefont {Huntley}},
  \bibinfo {author} {\bibfnamefont {N.~C.}\ \bibnamefont {Durand}}, \bibinfo
  {author} {\bibfnamefont {E.~K.}\ \bibnamefont {Stamenova}}, \bibinfo {author}
  {\bibfnamefont {I.~D.}\ \bibnamefont {Bochkov}}, \bibinfo {author}
  {\bibfnamefont {J.~T.}\ \bibnamefont {Robinson}}, \bibinfo {author}
  {\bibfnamefont {A.~L.}\ \bibnamefont {Sanborn}}, \bibinfo {author}
  {\bibfnamefont {I.}~\bibnamefont {Machol}}, \bibinfo {author} {\bibfnamefont
  {A.~D.}\ \bibnamefont {Omer}}, \bibinfo {author} {\bibfnamefont {E.~S.}\
  \bibnamefont {Lander}}, \emph {et~al.},\ }\href@noop {} {\bibfield  {journal}
  {\bibinfo  {journal} {Cell}\ }\textbf {\bibinfo {volume} {159}},\ \bibinfo
  {pages} {1665} (\bibinfo {year} {2014})}\BibitemShut {NoStop}%
\bibitem [{\citenamefont {Wang}\ \emph {et~al.}(2016)\citenamefont {Wang},
  \citenamefont {Su}, \citenamefont {Beliveau}, \citenamefont {Bintu},
  \citenamefont {Moffitt}, \citenamefont {Wu},\ and\ \citenamefont
  {Zhuang}}]{wang2016spatial}%
  \BibitemOpen
  \bibfield  {author} {\bibinfo {author} {\bibfnamefont {S.}~\bibnamefont
  {Wang}}, \bibinfo {author} {\bibfnamefont {J.-H.}\ \bibnamefont {Su}},
  \bibinfo {author} {\bibfnamefont {B.~J.}\ \bibnamefont {Beliveau}}, \bibinfo
  {author} {\bibfnamefont {B.}~\bibnamefont {Bintu}}, \bibinfo {author}
  {\bibfnamefont {J.~R.}\ \bibnamefont {Moffitt}}, \bibinfo {author}
  {\bibfnamefont {C.-t.}\ \bibnamefont {Wu}},\ and\ \bibinfo {author}
  {\bibfnamefont {X.}~\bibnamefont {Zhuang}},\ }\href@noop {} {\bibfield
  {journal} {\bibinfo  {journal} {Science}\ }\textbf {\bibinfo {volume}
  {353}},\ \bibinfo {pages} {598} (\bibinfo {year} {2016})}\BibitemShut
  {NoStop}%
\bibitem [{\citenamefont {Bintu}\ \emph {et~al.}(2018)\citenamefont {Bintu},
  \citenamefont {Mateo}, \citenamefont {Su}, \citenamefont {Sinnott-Armstrong},
  \citenamefont {Parker}, \citenamefont {Kinrot}, \citenamefont {Yamaya},
  \citenamefont {Boettiger},\ and\ \citenamefont {Zhuang}}]{bintu2018super}%
  \BibitemOpen
  \bibfield  {author} {\bibinfo {author} {\bibfnamefont {B.}~\bibnamefont
  {Bintu}}, \bibinfo {author} {\bibfnamefont {L.~J.}\ \bibnamefont {Mateo}},
  \bibinfo {author} {\bibfnamefont {J.-H.}\ \bibnamefont {Su}}, \bibinfo
  {author} {\bibfnamefont {N.~A.}\ \bibnamefont {Sinnott-Armstrong}}, \bibinfo
  {author} {\bibfnamefont {M.}~\bibnamefont {Parker}}, \bibinfo {author}
  {\bibfnamefont {S.}~\bibnamefont {Kinrot}}, \bibinfo {author} {\bibfnamefont
  {K.}~\bibnamefont {Yamaya}}, \bibinfo {author} {\bibfnamefont {A.~N.}\
  \bibnamefont {Boettiger}},\ and\ \bibinfo {author} {\bibfnamefont
  {X.}~\bibnamefont {Zhuang}},\ }\href@noop {} {\bibfield  {journal} {\bibinfo
  {journal} {Science}\ }\textbf {\bibinfo {volume} {362}},\ \bibinfo {pages}
  {eaau1783} (\bibinfo {year} {2018})}\BibitemShut {NoStop}%
\bibitem [{\citenamefont {Takei}\ \emph {et~al.}(2021)\citenamefont {Takei},
  \citenamefont {Yun}, \citenamefont {Zheng}, \citenamefont {Ollikainen},
  \citenamefont {Pierson}, \citenamefont {White}, \citenamefont {Shah},
  \citenamefont {Thomassie}, \citenamefont {Suo}, \citenamefont {Eng} \emph
  {et~al.}}]{takei2021integrated}%
  \BibitemOpen
  \bibfield  {author} {\bibinfo {author} {\bibfnamefont {Y.}~\bibnamefont
  {Takei}}, \bibinfo {author} {\bibfnamefont {J.}~\bibnamefont {Yun}}, \bibinfo
  {author} {\bibfnamefont {S.}~\bibnamefont {Zheng}}, \bibinfo {author}
  {\bibfnamefont {N.}~\bibnamefont {Ollikainen}}, \bibinfo {author}
  {\bibfnamefont {N.}~\bibnamefont {Pierson}}, \bibinfo {author} {\bibfnamefont
  {J.}~\bibnamefont {White}}, \bibinfo {author} {\bibfnamefont
  {S.}~\bibnamefont {Shah}}, \bibinfo {author} {\bibfnamefont {J.}~\bibnamefont
  {Thomassie}}, \bibinfo {author} {\bibfnamefont {S.}~\bibnamefont {Suo}},
  \bibinfo {author} {\bibfnamefont {C.-H.~L.}\ \bibnamefont {Eng}}, \emph
  {et~al.},\ }\href@noop {} {\bibfield  {journal} {\bibinfo  {journal}
  {Nature}\ }\textbf {\bibinfo {volume} {590}},\ \bibinfo {pages} {344}
  (\bibinfo {year} {2021})}\BibitemShut {NoStop}%
\bibitem [{\citenamefont {Takaki}\ \emph {et~al.}(2025)\citenamefont {Takaki},
  \citenamefont {Savich}, \citenamefont {Brugu{\'e}s},\ and\ \citenamefont
  {J{\"u}licher}}]{takaki2025active}%
  \BibitemOpen
  \bibfield  {author} {\bibinfo {author} {\bibfnamefont {R.}~\bibnamefont
  {Takaki}}, \bibinfo {author} {\bibfnamefont {Y.}~\bibnamefont {Savich}},
  \bibinfo {author} {\bibfnamefont {J.}~\bibnamefont {Brugu{\'e}s}},\ and\
  \bibinfo {author} {\bibfnamefont {F.}~\bibnamefont {J{\"u}licher}},\
  }\href@noop {} {\bibfield  {journal} {\bibinfo  {journal} {Physical review
  letters}\ }\textbf {\bibinfo {volume} {134}},\ \bibinfo {pages} {128401}
  (\bibinfo {year} {2025})}\BibitemShut {NoStop}%
\bibitem [{\citenamefont {Boettiger}\ \emph {et~al.}(2016)\citenamefont
  {Boettiger}, \citenamefont {Bintu}, \citenamefont {Moffitt}, \citenamefont
  {Wang}, \citenamefont {Beliveau}, \citenamefont {Fudenberg}, \citenamefont
  {Imakaev}, \citenamefont {Mirny}, \citenamefont {Wu},\ and\ \citenamefont
  {Zhuang}}]{boettiger2016super}%
  \BibitemOpen
  \bibfield  {author} {\bibinfo {author} {\bibfnamefont {A.~N.}\ \bibnamefont
  {Boettiger}}, \bibinfo {author} {\bibfnamefont {B.}~\bibnamefont {Bintu}},
  \bibinfo {author} {\bibfnamefont {J.~R.}\ \bibnamefont {Moffitt}}, \bibinfo
  {author} {\bibfnamefont {S.}~\bibnamefont {Wang}}, \bibinfo {author}
  {\bibfnamefont {B.~J.}\ \bibnamefont {Beliveau}}, \bibinfo {author}
  {\bibfnamefont {G.}~\bibnamefont {Fudenberg}}, \bibinfo {author}
  {\bibfnamefont {M.}~\bibnamefont {Imakaev}}, \bibinfo {author} {\bibfnamefont
  {L.~A.}\ \bibnamefont {Mirny}}, \bibinfo {author} {\bibfnamefont {C.-t.}\
  \bibnamefont {Wu}},\ and\ \bibinfo {author} {\bibfnamefont {X.}~\bibnamefont
  {Zhuang}},\ }\href@noop {} {\bibfield  {journal} {\bibinfo  {journal}
  {Nature}\ }\textbf {\bibinfo {volume} {529}},\ \bibinfo {pages} {418}
  (\bibinfo {year} {2016})}\BibitemShut {NoStop}%
\bibitem [{\citenamefont {Rao}\ \emph {et~al.}(2017)\citenamefont {Rao},
  \citenamefont {Huang}, \citenamefont {St~Hilaire}, \citenamefont {Engreitz},
  \citenamefont {Perez}, \citenamefont {Kieffer-Kwon}, \citenamefont {Sanborn},
  \citenamefont {Johnstone}, \citenamefont {Bascom}, \citenamefont {Bochkov}
  \emph {et~al.}}]{rao2017cohesin}%
  \BibitemOpen
  \bibfield  {author} {\bibinfo {author} {\bibfnamefont {S.~S.}\ \bibnamefont
  {Rao}}, \bibinfo {author} {\bibfnamefont {S.-C.}\ \bibnamefont {Huang}},
  \bibinfo {author} {\bibfnamefont {B.~G.}\ \bibnamefont {St~Hilaire}},
  \bibinfo {author} {\bibfnamefont {J.~M.}\ \bibnamefont {Engreitz}}, \bibinfo
  {author} {\bibfnamefont {E.~M.}\ \bibnamefont {Perez}}, \bibinfo {author}
  {\bibfnamefont {K.-R.}\ \bibnamefont {Kieffer-Kwon}}, \bibinfo {author}
  {\bibfnamefont {A.~L.}\ \bibnamefont {Sanborn}}, \bibinfo {author}
  {\bibfnamefont {S.~E.}\ \bibnamefont {Johnstone}}, \bibinfo {author}
  {\bibfnamefont {G.~D.}\ \bibnamefont {Bascom}}, \bibinfo {author}
  {\bibfnamefont {I.~D.}\ \bibnamefont {Bochkov}}, \emph {et~al.},\ }\href@noop
  {} {\bibfield  {journal} {\bibinfo  {journal} {Cell}\ }\textbf {\bibinfo
  {volume} {171}},\ \bibinfo {pages} {305} (\bibinfo {year}
  {2017})}\BibitemShut {NoStop}%
\bibitem [{\citenamefont {Schwarzer}\ \emph {et~al.}(2017)\citenamefont
  {Schwarzer}, \citenamefont {Abdennur}, \citenamefont {Goloborodko},
  \citenamefont {Pekowska}, \citenamefont {Fudenberg}, \citenamefont {Loe-Mie},
  \citenamefont {Fonseca}, \citenamefont {Huber}, \citenamefont {Haering},
  \citenamefont {Mirny} \emph {et~al.}}]{schwarzer2017two}%
  \BibitemOpen
  \bibfield  {author} {\bibinfo {author} {\bibfnamefont {W.}~\bibnamefont
  {Schwarzer}}, \bibinfo {author} {\bibfnamefont {N.}~\bibnamefont {Abdennur}},
  \bibinfo {author} {\bibfnamefont {A.}~\bibnamefont {Goloborodko}}, \bibinfo
  {author} {\bibfnamefont {A.}~\bibnamefont {Pekowska}}, \bibinfo {author}
  {\bibfnamefont {G.}~\bibnamefont {Fudenberg}}, \bibinfo {author}
  {\bibfnamefont {Y.}~\bibnamefont {Loe-Mie}}, \bibinfo {author} {\bibfnamefont
  {N.~A.}\ \bibnamefont {Fonseca}}, \bibinfo {author} {\bibfnamefont
  {W.}~\bibnamefont {Huber}}, \bibinfo {author} {\bibfnamefont {C.~H.}\
  \bibnamefont {Haering}}, \bibinfo {author} {\bibfnamefont {L.}~\bibnamefont
  {Mirny}}, \emph {et~al.},\ }\href@noop {} {\bibfield  {journal} {\bibinfo
  {journal} {Nature}\ }\textbf {\bibinfo {volume} {551}},\ \bibinfo {pages}
  {51} (\bibinfo {year} {2017})}\BibitemShut {NoStop}%
\bibitem [{\citenamefont {Kim}\ \emph {et~al.}(2017)\citenamefont {Kim},
  \citenamefont {Liachko}, \citenamefont {Brickner}, \citenamefont {Cook},
  \citenamefont {Noble}, \citenamefont {Brickner}, \citenamefont {Shendure},\
  and\ \citenamefont {Dunham}}]{kim2017dynamic}%
  \BibitemOpen
  \bibfield  {author} {\bibinfo {author} {\bibfnamefont {S.}~\bibnamefont
  {Kim}}, \bibinfo {author} {\bibfnamefont {I.}~\bibnamefont {Liachko}},
  \bibinfo {author} {\bibfnamefont {D.~G.}\ \bibnamefont {Brickner}}, \bibinfo
  {author} {\bibfnamefont {K.}~\bibnamefont {Cook}}, \bibinfo {author}
  {\bibfnamefont {W.~S.}\ \bibnamefont {Noble}}, \bibinfo {author}
  {\bibfnamefont {J.~H.}\ \bibnamefont {Brickner}}, \bibinfo {author}
  {\bibfnamefont {J.}~\bibnamefont {Shendure}},\ and\ \bibinfo {author}
  {\bibfnamefont {M.~J.}\ \bibnamefont {Dunham}},\ }\href@noop {} {\bibfield
  {journal} {\bibinfo  {journal} {Elife}\ }\textbf {\bibinfo {volume} {6}},\
  \bibinfo {pages} {e23623} (\bibinfo {year} {2017})}\BibitemShut {NoStop}%
\bibitem [{\citenamefont {Hsieh}\ \emph {et~al.}(2022)\citenamefont {Hsieh},
  \citenamefont {Cattoglio}, \citenamefont {Slobodyanyuk}, \citenamefont
  {Hansen}, \citenamefont {Darzacq},\ and\ \citenamefont
  {Tjian}}]{hsieh2022enhancer}%
  \BibitemOpen
  \bibfield  {author} {\bibinfo {author} {\bibfnamefont {T.-H.~S.}\
  \bibnamefont {Hsieh}}, \bibinfo {author} {\bibfnamefont {C.}~\bibnamefont
  {Cattoglio}}, \bibinfo {author} {\bibfnamefont {E.}~\bibnamefont
  {Slobodyanyuk}}, \bibinfo {author} {\bibfnamefont {A.~S.}\ \bibnamefont
  {Hansen}}, \bibinfo {author} {\bibfnamefont {X.}~\bibnamefont {Darzacq}},\
  and\ \bibinfo {author} {\bibfnamefont {R.}~\bibnamefont {Tjian}},\
  }\href@noop {} {\bibfield  {journal} {\bibinfo  {journal} {Nature genetics}\
  }\textbf {\bibinfo {volume} {54}},\ \bibinfo {pages} {1919} (\bibinfo {year}
  {2022})}\BibitemShut {NoStop}%
\bibitem [{\citenamefont {De~Gennes}(1979)}]{de1979scaling}%
  \BibitemOpen
  \bibfield  {author} {\bibinfo {author} {\bibfnamefont {P.-G.}\ \bibnamefont
  {De~Gennes}},\ }\href@noop {} {\emph {\bibinfo {title} {Scaling concepts in
  polymer physics}}}\ (\bibinfo  {publisher} {Cornell university press},\
  \bibinfo {year} {1979})\BibitemShut {NoStop}%
\bibitem [{\citenamefont {Rubinstein}\ and\ \citenamefont
  {Colby}(2003)}]{rubinstein2003polymer}%
  \BibitemOpen
  \bibfield  {author} {\bibinfo {author} {\bibfnamefont {M.}~\bibnamefont
  {Rubinstein}}\ and\ \bibinfo {author} {\bibfnamefont {R.~H.}\ \bibnamefont
  {Colby}},\ }\href@noop {} {\emph {\bibinfo {title} {Polymer physics}}}\
  (\bibinfo  {publisher} {Oxford university press},\ \bibinfo {year}
  {2003})\BibitemShut {NoStop}%
\bibitem [{\citenamefont {Mirny}(2011)}]{mirny2011fractal}%
  \BibitemOpen
  \bibfield  {author} {\bibinfo {author} {\bibfnamefont {L.~A.}\ \bibnamefont
  {Mirny}},\ }\href@noop {} {\bibfield  {journal} {\bibinfo  {journal}
  {Chromosome Res.}\ }\textbf {\bibinfo {volume} {19}},\ \bibinfo {pages} {37}
  (\bibinfo {year} {2011})}\BibitemShut {NoStop}%
\bibitem [{\citenamefont {Grosberg}\ \emph {et~al.}(1993)\citenamefont
  {Grosberg}, \citenamefont {Rabin}, \citenamefont {Havlin},\ and\
  \citenamefont {Neer}}]{grosberg1993crumpled}%
  \BibitemOpen
  \bibfield  {author} {\bibinfo {author} {\bibfnamefont {A.}~\bibnamefont
  {Grosberg}}, \bibinfo {author} {\bibfnamefont {Y.}~\bibnamefont {Rabin}},
  \bibinfo {author} {\bibfnamefont {S.}~\bibnamefont {Havlin}},\ and\ \bibinfo
  {author} {\bibfnamefont {A.}~\bibnamefont {Neer}},\ }\href@noop {} {\bibfield
   {journal} {\bibinfo  {journal} {Europhys. Lett.}\ }\textbf {\bibinfo
  {volume} {23}},\ \bibinfo {pages} {373} (\bibinfo {year} {1993})}\BibitemShut
  {NoStop}%
\bibitem [{\citenamefont {Halsey}\ \emph {et~al.}(1986)\citenamefont {Halsey},
  \citenamefont {Jensen}, \citenamefont {Kadanoff}, \citenamefont {Procaccia},\
  and\ \citenamefont {Shraiman}}]{halsey1986fractal}%
  \BibitemOpen
  \bibfield  {author} {\bibinfo {author} {\bibfnamefont {T.~C.}\ \bibnamefont
  {Halsey}}, \bibinfo {author} {\bibfnamefont {M.~H.}\ \bibnamefont {Jensen}},
  \bibinfo {author} {\bibfnamefont {L.~P.}\ \bibnamefont {Kadanoff}}, \bibinfo
  {author} {\bibfnamefont {I.}~\bibnamefont {Procaccia}},\ and\ \bibinfo
  {author} {\bibfnamefont {B.~I.}\ \bibnamefont {Shraiman}},\ }\href@noop {}
  {\bibfield  {journal} {\bibinfo  {journal} {Phys. Rev. A}\ }\textbf {\bibinfo
  {volume} {33}},\ \bibinfo {pages} {1141} (\bibinfo {year}
  {1986})}\BibitemShut {NoStop}%
\bibitem [{\citenamefont {Ott}(2002)}]{ott2002chaos}%
  \BibitemOpen
  \bibfield  {author} {\bibinfo {author} {\bibfnamefont {E.}~\bibnamefont
  {Ott}},\ }\href@noop {} {\emph {\bibinfo {title} {Chaos in dynamical
  systems}}}\ (\bibinfo  {publisher} {Cambridge university press},\ \bibinfo
  {year} {2002})\BibitemShut {NoStop}%
\bibitem [{\citenamefont {Salat}\ \emph {et~al.}(2017)\citenamefont {Salat},
  \citenamefont {Murcio},\ and\ \citenamefont
  {Arcaute}}]{salat2017multifractal}%
  \BibitemOpen
  \bibfield  {author} {\bibinfo {author} {\bibfnamefont {H.}~\bibnamefont
  {Salat}}, \bibinfo {author} {\bibfnamefont {R.}~\bibnamefont {Murcio}},\ and\
  \bibinfo {author} {\bibfnamefont {E.}~\bibnamefont {Arcaute}},\ }\href@noop
  {} {\bibfield  {journal} {\bibinfo  {journal} {Physica A: Statistical
  Mechanics and its Applications}\ }\textbf {\bibinfo {volume} {473}},\
  \bibinfo {pages} {467} (\bibinfo {year} {2017})}\BibitemShut {NoStop}%
\bibitem [{\citenamefont {Sanborn}\ \emph {et~al.}(2015)\citenamefont
  {Sanborn}, \citenamefont {Rao}, \citenamefont {Huang}, \citenamefont
  {Durand}, \citenamefont {Huntley}, \citenamefont {Jewett}, \citenamefont
  {Bochkov}, \citenamefont {Chinnappan}, \citenamefont {Cutkosky},
  \citenamefont {Li} \emph {et~al.}}]{sanborn2015chromatin}%
  \BibitemOpen
  \bibfield  {author} {\bibinfo {author} {\bibfnamefont {A.~L.}\ \bibnamefont
  {Sanborn}}, \bibinfo {author} {\bibfnamefont {S.~S.}\ \bibnamefont {Rao}},
  \bibinfo {author} {\bibfnamefont {S.-C.}\ \bibnamefont {Huang}}, \bibinfo
  {author} {\bibfnamefont {N.~C.}\ \bibnamefont {Durand}}, \bibinfo {author}
  {\bibfnamefont {M.~H.}\ \bibnamefont {Huntley}}, \bibinfo {author}
  {\bibfnamefont {A.~I.}\ \bibnamefont {Jewett}}, \bibinfo {author}
  {\bibfnamefont {I.~D.}\ \bibnamefont {Bochkov}}, \bibinfo {author}
  {\bibfnamefont {D.}~\bibnamefont {Chinnappan}}, \bibinfo {author}
  {\bibfnamefont {A.}~\bibnamefont {Cutkosky}}, \bibinfo {author}
  {\bibfnamefont {J.}~\bibnamefont {Li}}, \emph {et~al.},\ }\href@noop {}
  {\bibfield  {journal} {\bibinfo  {journal} {Proc. Natl. Acad. Sci. U.S.A.}\
  }\textbf {\bibinfo {volume} {112}},\ \bibinfo {pages} {E6456} (\bibinfo
  {year} {2015})}\BibitemShut {NoStop}%
\bibitem [{\citenamefont {Chan}\ and\ \citenamefont
  {Rubinstein}(2024)}]{chan2024activity}%
  \BibitemOpen
  \bibfield  {author} {\bibinfo {author} {\bibfnamefont {B.}~\bibnamefont
  {Chan}}\ and\ \bibinfo {author} {\bibfnamefont {M.}~\bibnamefont
  {Rubinstein}},\ }\href@noop {} {\bibfield  {journal} {\bibinfo  {journal}
  {Proc. Natl. Acad. Sci. U.S.A.}\ }\textbf {\bibinfo {volume} {121}},\
  \bibinfo {pages} {e2401494121} (\bibinfo {year} {2024})}\BibitemShut
  {NoStop}%
\bibitem [{\citenamefont {Frisch}\ and\ \citenamefont
  {Parisi}(1980)}]{frisch1980fully}%
  \BibitemOpen
  \bibfield  {author} {\bibinfo {author} {\bibfnamefont {U.}~\bibnamefont
  {Frisch}}\ and\ \bibinfo {author} {\bibfnamefont {G.}~\bibnamefont
  {Parisi}},\ }\href@noop {} {\bibfield  {journal} {\bibinfo  {journal} {New
  York Academy of Sciences, Annals}\ }\textbf {\bibinfo {volume} {357}},\
  \bibinfo {pages} {359} (\bibinfo {year} {1980})}\BibitemShut {NoStop}%
\bibitem [{\citenamefont {Benzi}\ \emph {et~al.}(1984)\citenamefont {Benzi},
  \citenamefont {Paladin}, \citenamefont {Parisi},\ and\ \citenamefont
  {Vulpiani}}]{benzi1984multifractal}%
  \BibitemOpen
  \bibfield  {author} {\bibinfo {author} {\bibfnamefont {R.}~\bibnamefont
  {Benzi}}, \bibinfo {author} {\bibfnamefont {G.}~\bibnamefont {Paladin}},
  \bibinfo {author} {\bibfnamefont {G.}~\bibnamefont {Parisi}},\ and\ \bibinfo
  {author} {\bibfnamefont {A.}~\bibnamefont {Vulpiani}},\ }\href@noop {}
  {\bibfield  {journal} {\bibinfo  {journal} {Journal of Physics A:
  Mathematical and General}\ }\textbf {\bibinfo {volume} {17}},\ \bibinfo
  {pages} {3521} (\bibinfo {year} {1984})}\BibitemShut {NoStop}%
\bibitem [{\citenamefont {She}\ and\ \citenamefont
  {Leveque}(1994)}]{she1994universal}%
  \BibitemOpen
  \bibfield  {author} {\bibinfo {author} {\bibfnamefont {Z.-S.}\ \bibnamefont
  {She}}\ and\ \bibinfo {author} {\bibfnamefont {E.}~\bibnamefont {Leveque}},\
  }\href@noop {} {\bibfield  {journal} {\bibinfo  {journal} {Physical review
  letters}\ }\textbf {\bibinfo {volume} {72}},\ \bibinfo {pages} {336}
  (\bibinfo {year} {1994})}\BibitemShut {NoStop}%
\bibitem [{\citenamefont {Ba{\l}yga}\ and\ \citenamefont
  {Bourne}(1995)}]{balyga1995interpretation}%
  \BibitemOpen
  \bibfield  {author} {\bibinfo {author} {\bibfnamefont {J.}~\bibnamefont
  {Ba{\l}yga}}\ and\ \bibinfo {author} {\bibfnamefont {J.}~\bibnamefont
  {Bourne}},\ }\href@noop {} {\bibfield  {journal} {\bibinfo  {journal}
  {Chemical engineering science}\ }\textbf {\bibinfo {volume} {50}},\ \bibinfo
  {pages} {381} (\bibinfo {year} {1995})}\BibitemShut {NoStop}%
\bibitem [{\citenamefont {Paladin}\ and\ \citenamefont
  {Vulpiani}(1987{\natexlab{a}})}]{paladin1987degrees}%
  \BibitemOpen
  \bibfield  {author} {\bibinfo {author} {\bibfnamefont {G.}~\bibnamefont
  {Paladin}}\ and\ \bibinfo {author} {\bibfnamefont {A.}~\bibnamefont
  {Vulpiani}},\ }\href@noop {} {\bibfield  {journal} {\bibinfo  {journal}
  {Physical Review A}\ }\textbf {\bibinfo {volume} {35}},\ \bibinfo {pages}
  {1971} (\bibinfo {year} {1987}{\natexlab{a}})}\BibitemShut {NoStop}%
\bibitem [{\citenamefont {Paladin}\ and\ \citenamefont
  {Vulpiani}(1987{\natexlab{b}})}]{paladin1987anomalous}%
  \BibitemOpen
  \bibfield  {author} {\bibinfo {author} {\bibfnamefont {G.}~\bibnamefont
  {Paladin}}\ and\ \bibinfo {author} {\bibfnamefont {A.}~\bibnamefont
  {Vulpiani}},\ }\href@noop {} {\bibfield  {journal} {\bibinfo  {journal}
  {Physics Reports}\ }\textbf {\bibinfo {volume} {156}},\ \bibinfo {pages}
  {147} (\bibinfo {year} {1987}{\natexlab{b}})}\BibitemShut {NoStop}%
\bibitem [{\citenamefont {Chhabra}\ \emph {et~al.}(1989)\citenamefont
  {Chhabra}, \citenamefont {Meneveau}, \citenamefont {Jensen},\ and\
  \citenamefont {Sreenivasan}}]{chhabra1989direct}%
  \BibitemOpen
  \bibfield  {author} {\bibinfo {author} {\bibfnamefont {A.~B.}\ \bibnamefont
  {Chhabra}}, \bibinfo {author} {\bibfnamefont {C.}~\bibnamefont {Meneveau}},
  \bibinfo {author} {\bibfnamefont {R.~V.}\ \bibnamefont {Jensen}},\ and\
  \bibinfo {author} {\bibfnamefont {K.}~\bibnamefont {Sreenivasan}},\
  }\href@noop {} {\bibfield  {journal} {\bibinfo  {journal} {Physical Review
  A}\ }\textbf {\bibinfo {volume} {40}},\ \bibinfo {pages} {5284} (\bibinfo
  {year} {1989})}\BibitemShut {NoStop}%
\bibitem [{\citenamefont {Chhabra}\ and\ \citenamefont
  {Jensen}(1989)}]{chhabra1989direct2}%
  \BibitemOpen
  \bibfield  {author} {\bibinfo {author} {\bibfnamefont {A.}~\bibnamefont
  {Chhabra}}\ and\ \bibinfo {author} {\bibfnamefont {R.~V.}\ \bibnamefont
  {Jensen}},\ }\href@noop {} {\bibfield  {journal} {\bibinfo  {journal}
  {Physical Review Letters}\ }\textbf {\bibinfo {volume} {62}},\ \bibinfo
  {pages} {1327} (\bibinfo {year} {1989})}\BibitemShut {NoStop}%
\bibitem [{\citenamefont {Wink}\ \emph {et~al.}(2008)\citenamefont {Wink},
  \citenamefont {Bullmore}, \citenamefont {Barnes}, \citenamefont {Bernard},\
  and\ \citenamefont {Suckling}}]{wink2008monofractal}%
  \BibitemOpen
  \bibfield  {author} {\bibinfo {author} {\bibfnamefont {A.-M.}\ \bibnamefont
  {Wink}}, \bibinfo {author} {\bibfnamefont {E.}~\bibnamefont {Bullmore}},
  \bibinfo {author} {\bibfnamefont {A.}~\bibnamefont {Barnes}}, \bibinfo
  {author} {\bibfnamefont {F.}~\bibnamefont {Bernard}},\ and\ \bibinfo {author}
  {\bibfnamefont {J.}~\bibnamefont {Suckling}},\ }\href@noop {} {\bibfield
  {journal} {\bibinfo  {journal} {Human brain mapping}\ }\textbf {\bibinfo
  {volume} {29}},\ \bibinfo {pages} {791} (\bibinfo {year} {2008})}\BibitemShut
  {NoStop}%
\bibitem [{\citenamefont {Calvet}\ and\ \citenamefont
  {Fisher}(2002)}]{calvet2002multifractality}%
  \BibitemOpen
  \bibfield  {author} {\bibinfo {author} {\bibfnamefont {L.}~\bibnamefont
  {Calvet}}\ and\ \bibinfo {author} {\bibfnamefont {A.}~\bibnamefont
  {Fisher}},\ }\href@noop {} {\bibfield  {journal} {\bibinfo  {journal} {Review
  of Economics and Statistics}\ }\textbf {\bibinfo {volume} {84}},\ \bibinfo
  {pages} {381} (\bibinfo {year} {2002})}\BibitemShut {NoStop}%
\bibitem [{\citenamefont {Jiang}\ \emph {et~al.}(2019)\citenamefont {Jiang},
  \citenamefont {Xie}, \citenamefont {Zhou},\ and\ \citenamefont
  {Sornette}}]{jiang2019multifractal}%
  \BibitemOpen
  \bibfield  {author} {\bibinfo {author} {\bibfnamefont {Z.-Q.}\ \bibnamefont
  {Jiang}}, \bibinfo {author} {\bibfnamefont {W.-J.}\ \bibnamefont {Xie}},
  \bibinfo {author} {\bibfnamefont {W.-X.}\ \bibnamefont {Zhou}},\ and\
  \bibinfo {author} {\bibfnamefont {D.}~\bibnamefont {Sornette}},\ }\href@noop
  {} {\bibfield  {journal} {\bibinfo  {journal} {Reports on Progress in
  Physics}\ }\textbf {\bibinfo {volume} {82}},\ \bibinfo {pages} {125901}
  (\bibinfo {year} {2019})}\BibitemShut {NoStop}%
\bibitem [{\citenamefont {Jung}\ \emph {et~al.}(2020)\citenamefont {Jung},
  \citenamefont {Le}, \citenamefont {Mafwele}, \citenamefont {Lee},
  \citenamefont {Chae},\ and\ \citenamefont {Lee}}]{jung2020fractality}%
  \BibitemOpen
  \bibfield  {author} {\bibinfo {author} {\bibfnamefont {N.}~\bibnamefont
  {Jung}}, \bibinfo {author} {\bibfnamefont {Q.~A.}\ \bibnamefont {Le}},
  \bibinfo {author} {\bibfnamefont {B.~J.}\ \bibnamefont {Mafwele}}, \bibinfo
  {author} {\bibfnamefont {H.~M.}\ \bibnamefont {Lee}}, \bibinfo {author}
  {\bibfnamefont {S.~Y.}\ \bibnamefont {Chae}},\ and\ \bibinfo {author}
  {\bibfnamefont {J.~W.}\ \bibnamefont {Lee}},\ }\href@noop {} {\bibfield
  {journal} {\bibinfo  {journal} {Journal of the Korean Physical Society}\
  }\textbf {\bibinfo {volume} {77}},\ \bibinfo {pages} {186} (\bibinfo {year}
  {2020})}\BibitemShut {NoStop}%
\bibitem [{\citenamefont {Murcio}\ \emph {et~al.}(2015)\citenamefont {Murcio},
  \citenamefont {Masucci}, \citenamefont {Arcaute},\ and\ \citenamefont
  {Batty}}]{murcio2015multifractal}%
  \BibitemOpen
  \bibfield  {author} {\bibinfo {author} {\bibfnamefont {R.}~\bibnamefont
  {Murcio}}, \bibinfo {author} {\bibfnamefont {A.~P.}\ \bibnamefont {Masucci}},
  \bibinfo {author} {\bibfnamefont {E.}~\bibnamefont {Arcaute}},\ and\ \bibinfo
  {author} {\bibfnamefont {M.}~\bibnamefont {Batty}},\ }\href@noop {}
  {\bibfield  {journal} {\bibinfo  {journal} {Physical Review E}\ }\textbf
  {\bibinfo {volume} {92}},\ \bibinfo {pages} {062130} (\bibinfo {year}
  {2015})}\BibitemShut {NoStop}%
\bibitem [{\citenamefont {Liu}\ \emph {et~al.}(2017)\citenamefont {Liu},
  \citenamefont {Wang}, \citenamefont {Yu},\ and\ \citenamefont
  {Xie}}]{liu2017fractal}%
  \BibitemOpen
  \bibfield  {author} {\bibinfo {author} {\bibfnamefont {J.-L.}\ \bibnamefont
  {Liu}}, \bibinfo {author} {\bibfnamefont {J.}~\bibnamefont {Wang}}, \bibinfo
  {author} {\bibfnamefont {Z.-G.}\ \bibnamefont {Yu}},\ and\ \bibinfo {author}
  {\bibfnamefont {X.-H.}\ \bibnamefont {Xie}},\ }\href@noop {} {\bibfield
  {journal} {\bibinfo  {journal} {Scientific reports}\ }\textbf {\bibinfo
  {volume} {7}},\ \bibinfo {pages} {45588} (\bibinfo {year}
  {2017})}\BibitemShut {NoStop}%
\bibitem [{\citenamefont {Pigolotti}\ \emph {et~al.}(2020)\citenamefont
  {Pigolotti}, \citenamefont {Jensen}, \citenamefont {Zhan},\ and\
  \citenamefont {Tiana}}]{pigolotti2020bifractal}%
  \BibitemOpen
  \bibfield  {author} {\bibinfo {author} {\bibfnamefont {S.}~\bibnamefont
  {Pigolotti}}, \bibinfo {author} {\bibfnamefont {M.~H.}\ \bibnamefont
  {Jensen}}, \bibinfo {author} {\bibfnamefont {Y.}~\bibnamefont {Zhan}},\ and\
  \bibinfo {author} {\bibfnamefont {G.}~\bibnamefont {Tiana}},\ }\href@noop {}
  {\bibfield  {journal} {\bibinfo  {journal} {Phys. Rev. Res.}\ }\textbf
  {\bibinfo {volume} {2}},\ \bibinfo {pages} {043078} (\bibinfo {year}
  {2020})}\BibitemShut {NoStop}%
\bibitem [{\citenamefont {Lee}\ \emph {et~al.}(2025)\citenamefont {Lee},
  \citenamefont {Liu}, \citenamefont {Ziabkin},\ and\ \citenamefont
  {Zidovska}}]{LEE2025}%
  \BibitemOpen
  \bibfield  {author} {\bibinfo {author} {\bibfnamefont {S.}~\bibnamefont
  {Lee}}, \bibinfo {author} {\bibfnamefont {X.}~\bibnamefont {Liu}}, \bibinfo
  {author} {\bibfnamefont {I.}~\bibnamefont {Ziabkin}},\ and\ \bibinfo {author}
  {\bibfnamefont {A.}~\bibnamefont {Zidovska}},\ }\bibfield  {journal}
  {\bibinfo  {journal} {Biophysical Journal}\ }\href
  {https://doi.org/https://doi.org/10.1016/j.bpj.2025.02.014}
  {https://doi.org/10.1016/j.bpj.2025.02.014} (\bibinfo {year} {2025}),\
  \bibinfo {note} {in press}\BibitemShut {NoStop}%
\bibitem [{\citenamefont {Georges}\ \emph {et~al.}(2015)\citenamefont
  {Georges}, \citenamefont {Li}, \citenamefont {Lian}, \citenamefont
  {O'Meally}, \citenamefont {Deakin}, \citenamefont {Wang}, \citenamefont
  {Zhang}, \citenamefont {Fujita}, \citenamefont {Patel}, \citenamefont
  {Holleley} \emph {et~al.}}]{CentralBeardedDragon}%
  \BibitemOpen
  \bibfield  {author} {\bibinfo {author} {\bibfnamefont {A.}~\bibnamefont
  {Georges}}, \bibinfo {author} {\bibfnamefont {Q.}~\bibnamefont {Li}},
  \bibinfo {author} {\bibfnamefont {J.}~\bibnamefont {Lian}}, \bibinfo {author}
  {\bibfnamefont {D.}~\bibnamefont {O'Meally}}, \bibinfo {author}
  {\bibfnamefont {J.}~\bibnamefont {Deakin}}, \bibinfo {author} {\bibfnamefont
  {Z.}~\bibnamefont {Wang}}, \bibinfo {author} {\bibfnamefont {P.}~\bibnamefont
  {Zhang}}, \bibinfo {author} {\bibfnamefont {M.}~\bibnamefont {Fujita}},
  \bibinfo {author} {\bibfnamefont {H.~R.}\ \bibnamefont {Patel}}, \bibinfo
  {author} {\bibfnamefont {C.~E.}\ \bibnamefont {Holleley}}, \emph {et~al.},\
  }\href@noop {} {\bibfield  {journal} {\bibinfo  {journal} {Gigascience}\
  }\textbf {\bibinfo {volume} {4}},\ \bibinfo {pages} {s13742} (\bibinfo {year}
  {2015})}\BibitemShut {NoStop}%
\bibitem [{\citenamefont {Weber}\ \emph {et~al.}(2020)\citenamefont {Weber},
  \citenamefont {Park}, \citenamefont {Luria}, \citenamefont {Jeon},
  \citenamefont {Kim}, \citenamefont {Jeon}, \citenamefont {Bhak},
  \citenamefont {Jun}, \citenamefont {Kim}, \citenamefont {Hong} \emph
  {et~al.}}]{WhaleShark}%
  \BibitemOpen
  \bibfield  {author} {\bibinfo {author} {\bibfnamefont {J.~A.}\ \bibnamefont
  {Weber}}, \bibinfo {author} {\bibfnamefont {S.~G.}\ \bibnamefont {Park}},
  \bibinfo {author} {\bibfnamefont {V.}~\bibnamefont {Luria}}, \bibinfo
  {author} {\bibfnamefont {S.}~\bibnamefont {Jeon}}, \bibinfo {author}
  {\bibfnamefont {H.-M.}\ \bibnamefont {Kim}}, \bibinfo {author} {\bibfnamefont
  {Y.}~\bibnamefont {Jeon}}, \bibinfo {author} {\bibfnamefont {Y.}~\bibnamefont
  {Bhak}}, \bibinfo {author} {\bibfnamefont {J.~H.}\ \bibnamefont {Jun}},
  \bibinfo {author} {\bibfnamefont {S.~W.}\ \bibnamefont {Kim}}, \bibinfo
  {author} {\bibfnamefont {W.~H.}\ \bibnamefont {Hong}}, \emph {et~al.},\
  }\href@noop {} {\bibfield  {journal} {\bibinfo  {journal} {Proceedings of the
  National Academy of Sciences}\ }\textbf {\bibinfo {volume} {117}},\ \bibinfo
  {pages} {20662} (\bibinfo {year} {2020})}\BibitemShut {NoStop}%
\bibitem [{\citenamefont {Hirakawa}\ \emph {et~al.}(2016)\citenamefont
  {Hirakawa}, \citenamefont {Kaur}, \citenamefont {Shirasawa}, \citenamefont
  {Nichols}, \citenamefont {Nagano}, \citenamefont {Appels}, \citenamefont
  {Erskine},\ and\ \citenamefont {Isobe}}]{Clover}%
  \BibitemOpen
  \bibfield  {author} {\bibinfo {author} {\bibfnamefont {H.}~\bibnamefont
  {Hirakawa}}, \bibinfo {author} {\bibfnamefont {P.}~\bibnamefont {Kaur}},
  \bibinfo {author} {\bibfnamefont {K.}~\bibnamefont {Shirasawa}}, \bibinfo
  {author} {\bibfnamefont {P.}~\bibnamefont {Nichols}}, \bibinfo {author}
  {\bibfnamefont {S.}~\bibnamefont {Nagano}}, \bibinfo {author} {\bibfnamefont
  {R.}~\bibnamefont {Appels}}, \bibinfo {author} {\bibfnamefont
  {W.}~\bibnamefont {Erskine}},\ and\ \bibinfo {author} {\bibfnamefont {S.~N.}\
  \bibnamefont {Isobe}},\ }\href@noop {} {\bibfield  {journal} {\bibinfo
  {journal} {Scientific Reports}\ }\textbf {\bibinfo {volume} {6}},\ \bibinfo
  {pages} {30358} (\bibinfo {year} {2016})}\BibitemShut {NoStop}%
\bibitem [{WAG(2025)}]{WAGA}%
  \BibitemOpen
  \href@noop {} {\bibinfo {title} {{Western Australia Genome Atlas}}},\
  \bibinfo {howpublished} {\url{https://www.dnazoo.org/waga}} (\bibinfo {year}
  {2025}),\ \bibinfo {note} {accessed: 2025-11-14}\BibitemShut {NoStop}%
\bibitem [{\citenamefont {Dudchenko}\ \emph {et~al.}(2017)\citenamefont
  {Dudchenko}, \citenamefont {Batra}, \citenamefont {Omer}, \citenamefont
  {Nyquist}, \citenamefont {Hoeger}, \citenamefont {Durand}, \citenamefont
  {Shamim}, \citenamefont {Machol}, \citenamefont {Lander}, \citenamefont
  {Aiden} \emph {et~al.}}]{dudchenko2017novo}%
  \BibitemOpen
  \bibfield  {author} {\bibinfo {author} {\bibfnamefont {O.}~\bibnamefont
  {Dudchenko}}, \bibinfo {author} {\bibfnamefont {S.~S.}\ \bibnamefont
  {Batra}}, \bibinfo {author} {\bibfnamefont {A.~D.}\ \bibnamefont {Omer}},
  \bibinfo {author} {\bibfnamefont {S.~K.}\ \bibnamefont {Nyquist}}, \bibinfo
  {author} {\bibfnamefont {M.}~\bibnamefont {Hoeger}}, \bibinfo {author}
  {\bibfnamefont {N.~C.}\ \bibnamefont {Durand}}, \bibinfo {author}
  {\bibfnamefont {M.~S.}\ \bibnamefont {Shamim}}, \bibinfo {author}
  {\bibfnamefont {I.}~\bibnamefont {Machol}}, \bibinfo {author} {\bibfnamefont
  {E.~S.}\ \bibnamefont {Lander}}, \bibinfo {author} {\bibfnamefont {A.~P.}\
  \bibnamefont {Aiden}}, \emph {et~al.},\ }\href@noop {} {\bibfield  {journal}
  {\bibinfo  {journal} {Science}\ }\textbf {\bibinfo {volume} {356}},\ \bibinfo
  {pages} {92} (\bibinfo {year} {2017})}\BibitemShut {NoStop}%
\bibitem [{\citenamefont {Dudchenko}\ \emph {et~al.}(2018)\citenamefont
  {Dudchenko}, \citenamefont {Shamim}, \citenamefont {Batra}, \citenamefont
  {Durand}, \citenamefont {Musial}, \citenamefont {Mostofa}, \citenamefont
  {Pham}, \citenamefont {Glenn St~Hilaire}, \citenamefont {Yao}, \citenamefont
  {Stamenova}, \citenamefont {Hoeger}, \citenamefont {Nyquist}, \citenamefont
  {Korchina}, \citenamefont {Pletch}, \citenamefont {Flanagan}, \citenamefont
  {Tomaszewicz}, \citenamefont {McAloose}, \citenamefont {P{\'e}rez~Estrada},
  \citenamefont {Novak}, \citenamefont {Omer},\ and\ \citenamefont
  {Aiden}}]{Dudchenko254797}%
  \BibitemOpen
  \bibfield  {author} {\bibinfo {author} {\bibfnamefont {O.}~\bibnamefont
  {Dudchenko}}, \bibinfo {author} {\bibfnamefont {M.~S.}\ \bibnamefont
  {Shamim}}, \bibinfo {author} {\bibfnamefont {S.~S.}\ \bibnamefont {Batra}},
  \bibinfo {author} {\bibfnamefont {N.~C.}\ \bibnamefont {Durand}}, \bibinfo
  {author} {\bibfnamefont {N.~T.}\ \bibnamefont {Musial}}, \bibinfo {author}
  {\bibfnamefont {R.}~\bibnamefont {Mostofa}}, \bibinfo {author} {\bibfnamefont
  {M.}~\bibnamefont {Pham}}, \bibinfo {author} {\bibfnamefont {B.}~\bibnamefont
  {Glenn St~Hilaire}}, \bibinfo {author} {\bibfnamefont {W.}~\bibnamefont
  {Yao}}, \bibinfo {author} {\bibfnamefont {E.}~\bibnamefont {Stamenova}},
  \bibinfo {author} {\bibfnamefont {M.}~\bibnamefont {Hoeger}}, \bibinfo
  {author} {\bibfnamefont {S.~K.}\ \bibnamefont {Nyquist}}, \bibinfo {author}
  {\bibfnamefont {V.}~\bibnamefont {Korchina}}, \bibinfo {author}
  {\bibfnamefont {K.}~\bibnamefont {Pletch}}, \bibinfo {author} {\bibfnamefont
  {J.~P.}\ \bibnamefont {Flanagan}}, \bibinfo {author} {\bibfnamefont
  {A.}~\bibnamefont {Tomaszewicz}}, \bibinfo {author} {\bibfnamefont
  {D.}~\bibnamefont {McAloose}}, \bibinfo {author} {\bibfnamefont
  {C.}~\bibnamefont {P{\'e}rez~Estrada}}, \bibinfo {author} {\bibfnamefont
  {B.~J.}\ \bibnamefont {Novak}}, \bibinfo {author} {\bibfnamefont {A.~D.}\
  \bibnamefont {Omer}},\ and\ \bibinfo {author} {\bibfnamefont {E.~L.}\
  \bibnamefont {Aiden}},\ }\href {https://doi.org/10.1101/254797} {\bibfield
  {journal} {\bibinfo  {journal} {bioRxiv}\ ,\ \bibinfo {pages} {254797}}
  (\bibinfo {year} {2018})}\BibitemShut {NoStop}%
\bibitem [{\citenamefont {Hoencamp}\ \emph {et~al.}(2021)\citenamefont
  {Hoencamp}, \citenamefont {Dudchenko}, \citenamefont {Elbatsh}, \citenamefont
  {Brahmachari}, \citenamefont {Raaijmakers}, \citenamefont {van Schaik},
  \citenamefont {Sede{\~n}o~Cacciatore}, \citenamefont {Contessoto},
  \citenamefont {van Heesbeen}, \citenamefont {van~den Broek} \emph
  {et~al.}}]{hoencamp20213d}%
  \BibitemOpen
  \bibfield  {author} {\bibinfo {author} {\bibfnamefont {C.}~\bibnamefont
  {Hoencamp}}, \bibinfo {author} {\bibfnamefont {O.}~\bibnamefont {Dudchenko}},
  \bibinfo {author} {\bibfnamefont {A.~M.}\ \bibnamefont {Elbatsh}}, \bibinfo
  {author} {\bibfnamefont {S.}~\bibnamefont {Brahmachari}}, \bibinfo {author}
  {\bibfnamefont {J.~A.}\ \bibnamefont {Raaijmakers}}, \bibinfo {author}
  {\bibfnamefont {T.}~\bibnamefont {van Schaik}}, \bibinfo {author}
  {\bibfnamefont {{\'A}.}~\bibnamefont {Sede{\~n}o~Cacciatore}}, \bibinfo
  {author} {\bibfnamefont {V.~G.}\ \bibnamefont {Contessoto}}, \bibinfo
  {author} {\bibfnamefont {R.~G.}\ \bibnamefont {van Heesbeen}}, \bibinfo
  {author} {\bibfnamefont {B.}~\bibnamefont {van~den Broek}}, \emph {et~al.},\
  }\href@noop {} {\bibfield  {journal} {\bibinfo  {journal} {Science}\ }\textbf
  {\bibinfo {volume} {372}},\ \bibinfo {pages} {984} (\bibinfo {year}
  {2021})}\BibitemShut {NoStop}%
\bibitem [{\citenamefont {G{\"u}rsoy}\ \emph {et~al.}(2014)\citenamefont
  {G{\"u}rsoy}, \citenamefont {Xu}, \citenamefont {Kenter},\ and\ \citenamefont
  {Liang}}]{gursoy2014spatial}%
  \BibitemOpen
  \bibfield  {author} {\bibinfo {author} {\bibfnamefont {G.}~\bibnamefont
  {G{\"u}rsoy}}, \bibinfo {author} {\bibfnamefont {Y.}~\bibnamefont {Xu}},
  \bibinfo {author} {\bibfnamefont {A.~L.}\ \bibnamefont {Kenter}},\ and\
  \bibinfo {author} {\bibfnamefont {J.}~\bibnamefont {Liang}},\ }\href@noop {}
  {\bibfield  {journal} {\bibinfo  {journal} {Nucleic acids research}\ }\textbf
  {\bibinfo {volume} {42}},\ \bibinfo {pages} {8223} (\bibinfo {year}
  {2014})}\BibitemShut {NoStop}%
\bibitem [{\citenamefont {Chan}\ and\ \citenamefont
  {Rubinstein}(2023)}]{chan2023theory}%
  \BibitemOpen
  \bibfield  {author} {\bibinfo {author} {\bibfnamefont {B.}~\bibnamefont
  {Chan}}\ and\ \bibinfo {author} {\bibfnamefont {M.}~\bibnamefont
  {Rubinstein}},\ }\href@noop {} {\bibfield  {journal} {\bibinfo  {journal}
  {Proc. Natl. Acad. Sci. U.S.A.}\ }\textbf {\bibinfo {volume} {120}},\
  \bibinfo {pages} {e2222078120} (\bibinfo {year} {2023})}\BibitemShut
  {NoStop}%
\bibitem [{\citenamefont {Nagano}\ \emph {et~al.}(2013)\citenamefont {Nagano},
  \citenamefont {Lubling}, \citenamefont {Stevens}, \citenamefont
  {Schoenfelder}, \citenamefont {Yaffe}, \citenamefont {Dean}, \citenamefont
  {Laue}, \citenamefont {Tanay},\ and\ \citenamefont
  {Fraser}}]{nagano2013single}%
  \BibitemOpen
  \bibfield  {author} {\bibinfo {author} {\bibfnamefont {T.}~\bibnamefont
  {Nagano}}, \bibinfo {author} {\bibfnamefont {Y.}~\bibnamefont {Lubling}},
  \bibinfo {author} {\bibfnamefont {T.~J.}\ \bibnamefont {Stevens}}, \bibinfo
  {author} {\bibfnamefont {S.}~\bibnamefont {Schoenfelder}}, \bibinfo {author}
  {\bibfnamefont {E.}~\bibnamefont {Yaffe}}, \bibinfo {author} {\bibfnamefont
  {W.}~\bibnamefont {Dean}}, \bibinfo {author} {\bibfnamefont {E.~D.}\
  \bibnamefont {Laue}}, \bibinfo {author} {\bibfnamefont {A.}~\bibnamefont
  {Tanay}},\ and\ \bibinfo {author} {\bibfnamefont {P.}~\bibnamefont
  {Fraser}},\ }\href@noop {} {\bibfield  {journal} {\bibinfo  {journal}
  {Nature}\ }\textbf {\bibinfo {volume} {502}},\ \bibinfo {pages} {59}
  (\bibinfo {year} {2013})}\BibitemShut {NoStop}%
\bibitem [{\citenamefont {Halverson}\ \emph {et~al.}(2011)\citenamefont
  {Halverson}, \citenamefont {Lee}, \citenamefont {Grest}, \citenamefont
  {Grosberg},\ and\ \citenamefont {Kremer}}]{halverson2011molecular}%
  \BibitemOpen
  \bibfield  {author} {\bibinfo {author} {\bibfnamefont {J.~D.}\ \bibnamefont
  {Halverson}}, \bibinfo {author} {\bibfnamefont {W.~B.}\ \bibnamefont {Lee}},
  \bibinfo {author} {\bibfnamefont {G.~S.}\ \bibnamefont {Grest}}, \bibinfo
  {author} {\bibfnamefont {A.~Y.}\ \bibnamefont {Grosberg}},\ and\ \bibinfo
  {author} {\bibfnamefont {K.}~\bibnamefont {Kremer}},\ }\href@noop {}
  {\bibfield  {journal} {\bibinfo  {journal} {The Journal of chemical physics}\
  }\textbf {\bibinfo {volume} {134}} (\bibinfo {year} {2011})}\BibitemShut
  {NoStop}%
\bibitem [{\citenamefont {Grosberg}(2014)}]{grosberg2014annealed}%
  \BibitemOpen
  \bibfield  {author} {\bibinfo {author} {\bibfnamefont {A.~Y.}\ \bibnamefont
  {Grosberg}},\ }\href@noop {} {\bibfield  {journal} {\bibinfo  {journal} {Soft
  Matter}\ }\textbf {\bibinfo {volume} {10}},\ \bibinfo {pages} {560} (\bibinfo
  {year} {2014})}\BibitemShut {NoStop}%
\bibitem [{\citenamefont {Rosa}\ and\ \citenamefont
  {Everaers}(2017)}]{rosa2017beyond}%
  \BibitemOpen
  \bibfield  {author} {\bibinfo {author} {\bibfnamefont {A.}~\bibnamefont
  {Rosa}}\ and\ \bibinfo {author} {\bibfnamefont {R.}~\bibnamefont
  {Everaers}},\ }\href@noop {} {\bibfield  {journal} {\bibinfo  {journal}
  {Physical Review E}\ }\textbf {\bibinfo {volume} {95}},\ \bibinfo {pages}
  {012117} (\bibinfo {year} {2017})}\BibitemShut {NoStop}%
\bibitem [{\citenamefont {Smrek}\ and\ \citenamefont
  {Grosberg}(2013)}]{smrek2013novel}%
  \BibitemOpen
  \bibfield  {author} {\bibinfo {author} {\bibfnamefont {J.}~\bibnamefont
  {Smrek}}\ and\ \bibinfo {author} {\bibfnamefont {A.~Y.}\ \bibnamefont
  {Grosberg}},\ }\href@noop {} {\bibfield  {journal} {\bibinfo  {journal}
  {Physica A: Statistical Mechanics and its Applications}\ }\textbf {\bibinfo
  {volume} {392}},\ \bibinfo {pages} {6375} (\bibinfo {year}
  {2013})}\BibitemShut {NoStop}%
\bibitem [{\citenamefont {Halverson}\ \emph {et~al.}(2014)\citenamefont
  {Halverson}, \citenamefont {Smrek}, \citenamefont {Kremer},\ and\
  \citenamefont {Grosberg}}]{halverson2014melt}%
  \BibitemOpen
  \bibfield  {author} {\bibinfo {author} {\bibfnamefont {J.~D.}\ \bibnamefont
  {Halverson}}, \bibinfo {author} {\bibfnamefont {J.}~\bibnamefont {Smrek}},
  \bibinfo {author} {\bibfnamefont {K.}~\bibnamefont {Kremer}},\ and\ \bibinfo
  {author} {\bibfnamefont {A.~Y.}\ \bibnamefont {Grosberg}},\ }\href@noop {}
  {\bibfield  {journal} {\bibinfo  {journal} {Rep. Prog. Phys.}\ }\textbf
  {\bibinfo {volume} {77}},\ \bibinfo {pages} {022601} (\bibinfo {year}
  {2014})}\BibitemShut {NoStop}%
\bibitem [{\citenamefont {Rosa}\ and\ \citenamefont
  {Everaers}(2019)}]{rosa2019conformational}%
  \BibitemOpen
  \bibfield  {author} {\bibinfo {author} {\bibfnamefont {A.}~\bibnamefont
  {Rosa}}\ and\ \bibinfo {author} {\bibfnamefont {R.}~\bibnamefont
  {Everaers}},\ }\href@noop {} {\bibfield  {journal} {\bibinfo  {journal} {The
  European Physical Journal E}\ }\textbf {\bibinfo {volume} {42}},\ \bibinfo
  {pages} {7} (\bibinfo {year} {2019})}\BibitemShut {NoStop}%
\bibitem [{\citenamefont {Fudenberg}\ \emph {et~al.}(2016)\citenamefont
  {Fudenberg}, \citenamefont {Imakaev}, \citenamefont {Lu}, \citenamefont
  {Goloborodko}, \citenamefont {Abdennur},\ and\ \citenamefont
  {Mirny}}]{fudenberg2016formation}%
  \BibitemOpen
  \bibfield  {author} {\bibinfo {author} {\bibfnamefont {G.}~\bibnamefont
  {Fudenberg}}, \bibinfo {author} {\bibfnamefont {M.}~\bibnamefont {Imakaev}},
  \bibinfo {author} {\bibfnamefont {C.}~\bibnamefont {Lu}}, \bibinfo {author}
  {\bibfnamefont {A.}~\bibnamefont {Goloborodko}}, \bibinfo {author}
  {\bibfnamefont {N.}~\bibnamefont {Abdennur}},\ and\ \bibinfo {author}
  {\bibfnamefont {L.~A.}\ \bibnamefont {Mirny}},\ }\href@noop {} {\bibfield
  {journal} {\bibinfo  {journal} {Cell reports}\ }\textbf {\bibinfo {volume}
  {15}},\ \bibinfo {pages} {2038} (\bibinfo {year} {2016})}\BibitemShut
  {NoStop}%
\bibitem [{\citenamefont {Kantelhardt}\ \emph {et~al.}(2002)\citenamefont
  {Kantelhardt}, \citenamefont {Zschiegner}, \citenamefont {Koscielny-Bunde},
  \citenamefont {Havlin}, \citenamefont {Bunde},\ and\ \citenamefont
  {Stanley}}]{kantelhardt2002multifractal}%
  \BibitemOpen
  \bibfield  {author} {\bibinfo {author} {\bibfnamefont {J.~W.}\ \bibnamefont
  {Kantelhardt}}, \bibinfo {author} {\bibfnamefont {S.~A.}\ \bibnamefont
  {Zschiegner}}, \bibinfo {author} {\bibfnamefont {E.}~\bibnamefont
  {Koscielny-Bunde}}, \bibinfo {author} {\bibfnamefont {S.}~\bibnamefont
  {Havlin}}, \bibinfo {author} {\bibfnamefont {A.}~\bibnamefont {Bunde}},\ and\
  \bibinfo {author} {\bibfnamefont {H.~E.}\ \bibnamefont {Stanley}},\
  }\href@noop {} {\bibfield  {journal} {\bibinfo  {journal} {Physica A:
  Statistical Mechanics and its Applications}\ }\textbf {\bibinfo {volume}
  {316}},\ \bibinfo {pages} {87} (\bibinfo {year} {2002})}\BibitemShut
  {NoStop}%
\bibitem [{\citenamefont {Gibcus}\ and\ \citenamefont
  {Dekker}(2013)}]{gibcus2013hierarchy}%
  \BibitemOpen
  \bibfield  {author} {\bibinfo {author} {\bibfnamefont {J.~H.}\ \bibnamefont
  {Gibcus}}\ and\ \bibinfo {author} {\bibfnamefont {J.}~\bibnamefont
  {Dekker}},\ }\href@noop {} {\bibfield  {journal} {\bibinfo  {journal}
  {Molecular cell}\ }\textbf {\bibinfo {volume} {49}},\ \bibinfo {pages} {773}
  (\bibinfo {year} {2013})}\BibitemShut {NoStop}%
\bibitem [{\citenamefont {Bernenko}\ \emph {et~al.}(2023)\citenamefont
  {Bernenko}, \citenamefont {Lee}, \citenamefont {Stenberg},\ and\
  \citenamefont {Lizana}}]{bernenko2023mapping}%
  \BibitemOpen
  \bibfield  {author} {\bibinfo {author} {\bibfnamefont {D.}~\bibnamefont
  {Bernenko}}, \bibinfo {author} {\bibfnamefont {S.~H.}\ \bibnamefont {Lee}},
  \bibinfo {author} {\bibfnamefont {P.}~\bibnamefont {Stenberg}},\ and\
  \bibinfo {author} {\bibfnamefont {L.}~\bibnamefont {Lizana}},\ }\href@noop {}
  {\bibfield  {journal} {\bibinfo  {journal} {PLOS Computational Biology}\
  }\textbf {\bibinfo {volume} {19}},\ \bibinfo {pages} {e1011185} (\bibinfo
  {year} {2023})}\BibitemShut {NoStop}%
\bibitem [{\citenamefont {Knight}\ and\ \citenamefont
  {Ruiz}(2013)}]{knight2013fast}%
  \BibitemOpen
  \bibfield  {author} {\bibinfo {author} {\bibfnamefont {P.~A.}\ \bibnamefont
  {Knight}}\ and\ \bibinfo {author} {\bibfnamefont {D.}~\bibnamefont {Ruiz}},\
  }\href@noop {} {\bibfield  {journal} {\bibinfo  {journal} {IMA Journal of
  Numerical Analysis}\ }\textbf {\bibinfo {volume} {33}},\ \bibinfo {pages}
  {1029} (\bibinfo {year} {2013})}\BibitemShut {NoStop}%
\bibitem [{\citenamefont {Durand}\ \emph {et~al.}(2016)\citenamefont {Durand},
  \citenamefont {Robinson}, \citenamefont {Shamim}, \citenamefont {Machol},
  \citenamefont {Mesirov}, \citenamefont {Lander},\ and\ \citenamefont
  {Aiden}}]{DURAND201699}%
  \BibitemOpen
  \bibfield  {author} {\bibinfo {author} {\bibfnamefont {N.~C.}\ \bibnamefont
  {Durand}}, \bibinfo {author} {\bibfnamefont {J.~T.}\ \bibnamefont
  {Robinson}}, \bibinfo {author} {\bibfnamefont {M.~S.}\ \bibnamefont
  {Shamim}}, \bibinfo {author} {\bibfnamefont {I.}~\bibnamefont {Machol}},
  \bibinfo {author} {\bibfnamefont {J.~P.}\ \bibnamefont {Mesirov}}, \bibinfo
  {author} {\bibfnamefont {E.~S.}\ \bibnamefont {Lander}},\ and\ \bibinfo
  {author} {\bibfnamefont {E.~L.}\ \bibnamefont {Aiden}},\ }\href
  {https://doi.org/https://doi.org/10.1016/j.cels.2015.07.012} {\bibfield
  {journal} {\bibinfo  {journal} {Cell Systems}\ }\textbf {\bibinfo {volume}
  {3}},\ \bibinfo {pages} {99} (\bibinfo {year} {2016})}\BibitemShut {NoStop}%
\bibitem [{\citenamefont {Falconer}(1994)}]{falconer1994multifractal}%
  \BibitemOpen
  \bibfield  {author} {\bibinfo {author} {\bibfnamefont {K.~J.}\ \bibnamefont
  {Falconer}},\ }\href@noop {} {\bibfield  {journal} {\bibinfo  {journal}
  {Journal of theoretical Probability}\ }\textbf {\bibinfo {volume} {7}},\
  \bibinfo {pages} {681} (\bibinfo {year} {1994})}\BibitemShut {NoStop}%
\bibitem [{\citenamefont {Fornberg}(1988)}]{fornberg1988generation}%
  \BibitemOpen
  \bibfield  {author} {\bibinfo {author} {\bibfnamefont {B.}~\bibnamefont
  {Fornberg}},\ }\href@noop {} {\bibfield  {journal} {\bibinfo  {journal}
  {Math. Comput.}\ }\textbf {\bibinfo {volume} {51}},\ \bibinfo {pages} {699}
  (\bibinfo {year} {1988})}\BibitemShut {NoStop}%
\bibitem [{our(2026)}]{our_git}%
  \BibitemOpen
  \href@noop {} {\bibinfo {title} {{All code for numerical calculations in this
  Git repository: [https://github.com/lizanalab/yang2026MF].}}} (\bibinfo
  {year} {2026})\BibitemShut {NoStop}%
\end{thebibliography}
%
%

%
\end{document}


\begin{CJK*}{UTF8}{}
\title{Supporting Information: Multifractal Scaling in Hi-C Maps}

\author{Seong-Gyu Yang \CJKfamily{mj}{(양성규)}}
\email{seong-gyu.yang@umu.se}
\affiliation{Integrated Science Lab, Department of Physics, Ume{\aa} University, SE-90187 Ume{\aa}, Sweden}
\affiliation{Department of Physics and Institute of Basic Science, Sungkyunkwan University, Suwon 16419, Republic of Korea}
    
\author{Lucas Hedstr\"om}
\affiliation{Department of Physics, Ume{\aa} University, SE-90187 Ume{\aa}, Sweden}

\author{Jan Smrek}
\affiliation{Faculty of Physics, University of Vienna, Boltzmanngasse 5, 1090 Vienna, Austria}

\author{Ludvig Lizana}
\email{ludvig.lizana@umu.se}
\affiliation{Integrated Science Lab, Department of Physics, Ume{\aa} University, SE-90187 Ume{\aa}, Sweden}
    
\date{\today}

\maketitle
\end{CJK*}
\section*{Further detailed derivation of $Z_\text{diag}(q,l)$ and $Z_\text{off}(q,l)$ for a single-exponent model}
In this Section, we provide the full derivation of the diagonal $Z_\text{diag}(q,l)$ and the off-diagonal $Z_\text{off}(q,l)$ contributions in the generalized partition function $Z(q,l)$ for a single-exponent model.

We begin by computing the mass $\mu_l(f=0)$ in a box at $f=0$.
From Eq.~5 in the main text together with the contact probability $P(s)$ in Eq.~2, we obtain
\begin{equation}\label{eq:single_mu_zero}
\begin{split}
\mu_l (f=0) &= \frac{l}{s_0} \int_0^{l} P(s)\, ds = \frac{l}{s_0} \left[ \int_0^{s_0} C\, ds + \int_{s_0}^l C \left( s/s_0 \right)^{-\gamma} ds  \right] =  C l + C l s_0^{\gamma-1} \int_{s_0}^l s^{-\gamma}\, ds\\
&= \begin{cases}
Cl \left[ 1 + \ln(l/s_0) \right], &\gamma = 1,\\
\frac{Cl}{1-\gamma} \left[ \left(l/s_0 \right)^{1-\gamma} - \gamma \right], &\gamma \neq 1,
\end{cases}
\ \approx \ 
\begin{cases}
\frac{\gamma C l}{\gamma-1} \sim l,&\gamma > 1,\\
Cl \left[ 1 + \ln(l/s_0) \right] \sim l, &\gamma = 1,\\
\frac{Cl}{1-\gamma} \left( l/s_0\right)^{1-\gamma} \sim l^{2-\gamma}, &\gamma < 1.
\end{cases}
\end{split}
\end{equation}
The mass $\mu_l(f=0)$ scales as $l$ when $\gamma\geq 1$ and as $l^{2-\gamma}$ when $\gamma < 1$.
The exponent of $\mu_l(f=0)$ is closely related to the bulk contact exponent $\beta_b$, which is referred as $\gamma_c$ in Ref.~\cite{rosa2017beyond}, because $\mu_l(f=0)$ counts the number of intra-segment contacts.

Substituting the results Eq.~\ref{eq:single_mu_zero} into Eq.~6 in the main text, we obtain the diagonal contribution $Z_\text{diag}(q,l)$ as
\begin{equation}\label{eq:single_detailed_Zdiag}
\begin{split}
Z_\text{diag}(q,l) &= 4\frac{L}{l} \left[ \mu_l (f=0)\right]^q 
=\begin{cases}
4 C^q L l^{q-1}\left[ 1 + \ln(l/s_0) \right]^q , &\gamma = 1,\\
\frac{4 C^q }{(1 - \gamma)^q} L l^{q-1}\left[ \left( l/s_0\right)^{1-\gamma} - \gamma \right]^q, &\gamma\neq 1,
\end{cases}\\
&\ \approx\ 
\begin{cases}
\frac{4 \gamma^q C^q }{(\gamma-1)^q} L l^{q-1} \sim l^{q-1} , &\gamma>1,\\
4 C^q L \left[ 1 + \ln(l/s_0) \right]^q l^{q-1} \sim l^{q-1}, &\gamma = 1,\\
\frac{4 C^q s_0^{q(\gamma-1)}}{(1 - \gamma)^q} L l^{q(2-\gamma)-1} \sim l^{q(2-\gamma)-1}, &\gamma<1.
\end{cases}
\end{split}
\end{equation}

For the off-diagonal contributions, we obtain the mass $\mu_l(f\neq 0)$ from Eq.~5 as
\begin{equation}\label{eq:single_mu_nonzero}
\begin{split}
\mu_l(f\neq0) &= \frac{l}{s_0} \int_{lf}^{l(f+1)} P(s)\, ds = \frac{l}{s_0} \int_{lf}^{l(f+1)} C\left( s/s_0 \right)^{-\gamma}\, ds = C l s_0^{\gamma-1} \int_{lf}^{l(f+1)} s^{-\gamma}\ ds\\
&=\begin{cases}
C l \left[ \ln(f+1) - \ln(f) \right], &\gamma=1,\\
\frac{C l }{1-\gamma} (l/s_0)^{1-\gamma} \left[(f+1)^{1-\gamma} - f^{1-\gamma} \right], & \gamma \neq 1,
\end{cases} \quad \approx \ C l (l/s_0)^{1-\gamma} f^{-\gamma}\sim l^{2-\gamma}.
\end{split}
\end{equation}
In the last line, we use approximations $(f+1)^{1-\gamma} - f^{1-\gamma} \approx (1-\gamma) f^{-\gamma}$ and $\ln(f+1) - \ln(f) \approx f^{-1}$.
Figure~\ref{fig:Approx} shows these approximations, and the difference decreases exponentially with $f$.
The scaling behavior of $\mu_l(f\neq0)$ is related to the surface contact exponent $\beta_s$, because the mass counts the inter-segment contacts.
The surface contact exponent $\beta_s$ is referred as surface exponent $\beta$ in Refs.~\cite{halverson2011molecular,smrek2013novel,halverson2014melt,rosa2017beyond}.

Next, we calculate $Z_\text{off}(q,l)$ using the result of Eq.~\ref{eq:single_mu_nonzero} as
\begin{equation}
\begin{split}
Z_\text{off}(q,l) &= 4  \int_1^{L/l} \left( L/l -f \right) \left[ \mu_l(f\neq 0)\right]^q df \\
&= 4 C^q L s_0^{q(\gamma-1)} l^{q(2-\gamma)-1} \int_1^{L/l} f^{-q\gamma} df \  - \ 4 C^q s_0^{q(\gamma-1)} l^{q(2-\gamma)} \int_1^{L/l} f^{1-q\gamma} df.
\end{split}
\end{equation}
Comparing these two terms, we find that the first term dominates by a factor of $L/l$.
We note that the box size $l$ lies in the regime $s_0 \ll l \ll L$.
The first term in $Z_\text{off}(q,l)$ yields
\begin{equation}
4C^q L s_0^{q(\gamma-1)} l^{q(2-\gamma)-1}\int_1^{L/l} f^{-q\gamma}df
=\begin{cases}
4 C^q L s_0^{1-q} l^{2(q-1)}\ln(L/l), &q\gamma=1,\\
\frac{4 C^q s_0^{q(\gamma-1)}}{1-q\gamma} L l^{q(2-\gamma)-1} \left[\left(L/l \right)^{1-q\gamma} -1 \right], &q\gamma\neq 1,
\end{cases}
\end{equation}
and the second term becomes
\begin{equation}
4 C^q s_0^{q(\gamma-1)} l^{q(2-\gamma)} \int_1^{L/l} f^{1-q\gamma}df
=\begin{cases}
4 C^q s_0^{2-q} l^{2(q-1)} \ln(L/l), & q\gamma =2,\\
\frac{4 C^q s_0^{q(\gamma-1)}}{2-q\gamma} l^{q(2-\gamma)} \left[\left(L/l\right)^{2-q\gamma} -1\right], &q\gamma \neq 2.
\end{cases}
\end{equation}
Thus, the off-diagonal term can be rewritten as
\begin{equation}\label{eq:single_detailed_Zoff}
\begin{split}
Z_\text{off}(q,l)
&=\begin{cases}
4 C^q s_0^{1-q} L l^{2(q-1)}\ln(L/l), &q\gamma=1,\\
\frac{4 C^q s_0^{q(\gamma-1)}}{1-q\gamma} L l^{q(2-\gamma)-1} \left[\left(L/l \right)^{1-q\gamma} -1 \right], &q\gamma\neq 1,
\end{cases}
\quad  -\quad  \begin{cases}
4 C^q s_0^{2-q} l^{2(q-1)} \ln(L/l), & q\gamma =2,\\
\frac{4 C^q s_0^{q(\gamma-1)}}{2-q\gamma} l^{q(2-\gamma)} \left[ \left(L/l\right)^{2-q\gamma} -1 \right], &q\gamma \neq 2,
\end{cases}\\
&\approx\begin{cases}
\frac{4 C^q s_0^{q(\gamma-1)}}{q\gamma-1} L l^{q(2-\gamma)-1} \sim l^{q(2-\gamma)-1}, &q\gamma > 1,\\
4 C^q  s_0^{1-q} L l^{2(q-1)}\ln(L/l) \sim l^{2(q-1)}, &q\gamma=1,\\
\frac{4 C^q s_0^{q(\gamma-1)}}{1-q\gamma} L^{2-q\gamma} l^{2(q-1)} \sim l^{2(q-1)}, &q\gamma < 1,
\end{cases}
\end{split}
\end{equation}
By only considering the dominant contributions in $Z_\text{diag}(q,l)$ and $Z_\text{off}(q,l)$, we obtain Eqs.~8 and 10 in the main text, which are plotted in Fig.~3A.

In Figs.~3B and 3C in the main text, the white dots represent the mass exponent $\tau(q)$ obtained from $Z(q,l) = Z_\text{diag}(q,l) + Z_\text{off}(q,l)$ using Eqs.~\ref{eq:single_detailed_Zdiag} and \ref{eq:single_detailed_Zoff}.

\section*{Detailed derivation of $Z_\text{diag}(q,l)$ and $Z_\text{off}(q,l)$ for a double-exponent model}
In a double-exponent model, the contact probability $P(s)$ follows Eq.~11 with a sharp transition of exponents from $\gamma$ to $\eta$ at $s_T$, with $\gamma<1$.
For the sake of generality, we allow $\eta$ to be either smaller or larger than unity.

\subsection*{For large box sizes ($l > s_T$)}
When the box size $l$ is larger than $s_T$, the results are qualitatively the same as in a single-exponent model, but the results have some additional terms and factors, and depend on $\eta$.
We note that we assume $s_0 \ll s_T < l \ll L$ for large box size.

In the same way as before, we calculate the mass $\mu_l(f=0)$ in a box at $f=0$, which yields
\begin{equation}\label{eq:double_mu_zero}
\begin{split}
\mu_l(f=0) &= \frac{l}{s_0} \int_0^lP(s)\, ds = \frac{l}{s_0}\left[ \int_0^{s_0}C\,ds + \int_{s_0}^{s_T}C\left(s/s_0\right)^{-\gamma}ds + \int_{s_T}^l C\left( s_T/s_0\right)^{-\gamma} \left( s/s_T \right)^{-\eta} ds\right]\\
&=C l + C l s_0^{\gamma-1}\int_{s_0}^{s_T} s^{-\gamma}\,ds + C l \left( s_T/s_0\right)^{1-\gamma}s_T^{\eta-1} \int_{s_T}^l s^{-\eta}\,ds\\
&= \begin{cases}
C l \left[  - \frac{\gamma}{1-\gamma} + \frac{1}{1-\gamma} \left( s_T/s_0\right)^{1-\gamma} + \left(s_T/s_0\right)^{1-\gamma}\ln(l/s_T) \right], &\eta=1,\\
C l \left[ -\frac{\gamma}{1-\gamma} - \frac{\eta-\gamma}{(1-\gamma)(1-\eta)} \left( s_T/s_0\right)^{1-\gamma} + \frac{1}{1-\eta}\left( s_T/s_0\right)^{1-\gamma}\left(l/s_T\right)^{1-\eta} \right], &\eta \neq 1,
\end{cases}\\
&\approx \begin{cases}
\frac{(\eta - \gamma) C l}{(1-\gamma)(\eta-1)} \left( s_T/s_0 \right)^{1-\gamma} \sim l, &\eta > 1,\\
\frac{Cl}{1-\gamma} \left(s_T/s_0\right)^{1-\gamma} \left[  1 + (1-\gamma)\ln(l/s_T) \right] \sim l, &\eta=1,\\
\frac{Cl}{1-\eta} \left( s_T/s_0\right)^{1-\gamma} \left( l/s_T\right)^{1-\eta} \sim l^{2-\eta}, &\eta < 1.
\end{cases}
\end{split}
\end{equation}
In this study, we assume that $\gamma <1$, thus we ignore the case when $\gamma\geq1$ in the second term.
The dominant term in the diagonal mass $\mu_l(f=0)$ scales as $l$ when $\eta\geq1$, and as $l^{2-\eta}$ when $\eta <1$, which matches the scaling behavior in Eq.~\ref{eq:single_mu_zero} upon replacing $\gamma$ with $\eta$.

Equation~\ref{eq:double_mu_zero} yields $Z_\text{diag}(q,l)$:
\begin{equation}\label{eq:double_detailed_Zdiag}
\begin{split}
Z_\text{diag}(q,l) &= 4 \frac{L}{l} \left[ \mu_l(f=0)\right]^q = 
\begin{cases}
\frac{4 C^q}{(1-\gamma)^q} L l^{q-1}\left[-\gamma + \left(s_T/s_0\right)^{1-\gamma} + \left( s_T/s_0\right)^{-\gamma}\ln(l/s_T)\right]^q, &\eta=1,\\
4 C^q L l^{q-1}\left[ \frac{\gamma}{\gamma-1} + \frac{\eta-\gamma}{(1-\gamma)(\eta-1)} \left( s_T/s_0\right)^{1-\gamma} + \frac{1}{1-\eta} \left(s_T/s_0\right)^{-\gamma}\left(l/s_T\right)^{1-\eta} \right]^q, &\eta \neq 1,
\end{cases}\\
&\approx
\begin{cases}
\frac{4(\eta-\gamma)^q C^q}{(1-\gamma)^q(\eta-1)^q} \left(s_T/s_0\right)^{q(1-\gamma)} L l^{q-1}\sim l^{q-1}, &\eta > 1,\\
\frac{4C^q}{(1-\gamma)^q} \left(s_T/s_0\right)^{q(1-\gamma)} L l^{q-1}\left[ 1 + (1-\gamma) \ln (l/s_T) \right] \sim l^{q-1}, &\eta = 1,\\
\frac{4 C^q s_T^{q(\eta-1)}}{(1-\eta)^q} \left(s_T/s_0\right)^{q(1-\gamma)} L l^{q(2-\eta)-1} \sim l^{q(2-\eta)-1}, &\eta < 1,
\end{cases}
\end{split}
\end{equation}
which exhibits the same scaling behavior in $l$, but replacing $\gamma$ with $\eta$ in Eq.~\ref{eq:single_detailed_Zdiag}.

The off-diagonal mass $\mu_l(f\neq 0)$ takes a form analogous to Eq.~\ref{eq:single_mu_nonzero}:
\begin{equation}\label{eq:double_mu_nonzero}
\begin{split}
\mu_l(f\neq 0) &= \frac{l}{s_0} \int_{lf}^{l(f+1)} P(s)ds = C l \left( s_T/s_0\right)^{1-\gamma} s_T^{\eta-1} \int_{lf}^{l(f+1)} s^{-\eta}ds \\
&\approx C l \left(s_T/s_0\right)^{1-\gamma} (l/s_T)^{1-\eta} f^{-\eta} \sim l^{2-\eta}.
\end{split}
\end{equation}
To obtain Eq.~\ref{eq:double_mu_nonzero}, we also use the same approximations that we used to calculate Eq.~\ref{eq:single_mu_nonzero}.
Using this result, the off-diagonal contribution $Z_\text{off}(q,l)$ is rewritten as
\begin{equation}\label{eq:double_detailed_Zoff}
\begin{split}
Z_\text{off}(q,l) &= 4 \int_1^{L/l} \left(L/l - f \right) \left[\mu_l(f\neq 0)\right]^q \,df\\
&= 4 C^q s_T^{q(\eta-1)} \left( s_T/s_0\right)^{q(1-\gamma)} L  l^{q(2-\eta)-1}\int_1^{L/l}f^{-q\eta}\,df \quad -\quad  4 C^q s_T^{q(\eta-1)} \left(s_T/s_0\right)^{q(1-\gamma)} l^{q(2-\eta)}\int_1^{L/l}f^{1-q\eta}\,df\\
&=
\begin{cases}
4C^q  s_T^{1-q} \left(s_T/s_0\right)^{q(1-\gamma)} L l ^{2(q-1)} \ln(L/l), &q\eta =1,\\
\frac{4C^q s_T^{q(\eta-1)}}{1-q\eta} \left(s_T/s_0\right)^{q(1-\gamma)} L l ^{q(2-\eta)-1}  \left[\left(L/l\right)^{1-q\eta} - 1\right], &q\eta \neq 1,
\end{cases}\\
&\quad - \begin{cases}
4 C^q s_T^{2-q} \left( s_T/s_0 \right)^{q(1-\gamma)} l^{2(q-2)} \ln(L/l), &q\eta = 2,\\
\frac{4 C^q s_T^{q(\eta-1)}}{2-q\eta}\left(s_T/s_0\right)^{q(1-\gamma)} l^{q(2-\eta)}\left[ \left(L/l\right)^{2-q\eta}-1\right], &q\eta \neq 2,
\end{cases}\\
&\approx
\begin{cases}
\frac{4C^q s_T^{q(\eta-1)}}{q\eta-1} \left( s_T/s_0 \right)^{q(1-\gamma)} L l^{q(2-\eta)-1} \sim l^{q(2-\eta)-1} , &q\eta>1,\\
4C^q  s_T^{1-q} \left(s_T/s_0\right)^{q(1-\gamma)} L l ^{2(q-1)} \ln(L/l)  \sim l^{2(q-1)}, &q\eta=1,\\
\frac{4C^q s_T^{q(\eta-1)}}{1-q\eta} \left( s_T/s_0 \right)^{q(1-\gamma)} L^{2-q\eta} l^{2(q-1)} \sim l^{2(q-1)} , &q\eta<1.
\end{cases}
\end{split}
\end{equation}
This result also exhibits the same scaling behavior as Eq.~\ref{eq:single_detailed_Zoff}, but with additional factors and replacing $\gamma$ with $\eta$.
Considering the dominant term in $Z(q,l) = Z_\text{diag}(q,l) + Z_\text{off}(q,l)$, Fig.~6C depicts the four regimes of $\tau(q)$.

\subsection*{For small box sizes ($l < s_T$)}

In the small-$l$ regime, we assume $s_0 \ll  l < s_T \ll L$.
In this case, the diagonal mass $\mu_l(f=0)$ has the same form as in Eq.~\ref{eq:single_mu_zero} for $\gamma <1$, and the resulting diagonal $Z_\text{diag}(q,l)$ follows Eq.~\ref{eq:single_detailed_Zdiag}.

The off-diagonal contribution, in contrast, differs from the previous results.
We distinguish three cases depending on the position of the box relative to the transition scale $s_T$:
the first, where the box covers only $s < s_T$ ($f < f_c$ with $f_c = \lfloor s_T/l \rfloor$), so that $P(s)$ decays only with $\gamma$;
the second, where the box covers only $s > s_T$ ($f > f_c$), so that $P(s)$ decays only with $\eta$;
and the third, where the box contains the transition point $s_T$ ($f = f_c$) thus both exponents contribute.
In the continuum approximation, the single point $f=f_c$ has zero measure, so we ignore its contribution.
Furthermore, since $l < s_T$ in the small-$l$ regime, we have $s_T/l > 1$ and approximate $f_c = \lfloor s_T/l\rfloor \approx s_T/l$.
The relative error of this approximation is $O(l/s_T)$, which does not affect the leading-order scaling.

Because $P(s)$ decays only with $\gamma$ in the first region ($f<f_c)$, the mass in this region $\mu_l(f<f_c)$ follows the same form as in Eq.~\ref{eq:single_mu_nonzero}, and in the second region ($f>f_c$), the mass $\mu_l(f>f_c)$ takes the same form as in Eq.~\ref{eq:double_mu_nonzero}.
Therefore, using Eqs.~\ref{eq:single_mu_nonzero} and \ref{eq:double_mu_nonzero}, we compute $Z_\text{off}(q,l)$ as
\begin{equation}\label{eq:double_detail_Zoff}
\begin{split}
Z_\text{off}(q,l) &= 4\int_1^{L/l}\left(L/l - f\right) \left[ \mu_l(f\neq0)\right]^qdf\\
&\approx 4 \int_1^{s_T/l} \left(L/l - f\right) \left[ \mu_l(f<f_c)\right]^qdf  \ +\   4\int_{s_T/l}^{L/l}\left(L/l - f\right) \left[ \mu_l(f>f_c)\right]^qdf\\
& = 4C^q l^q \left( l/s_0 \right)^{q(1-\gamma)} \int_1^{s_T/l}\left(L/l-f\right)f^{-q\gamma}df \ +\  4 C^q l^q \left(s_T/s_0\right)^{q(1-\gamma)} \left(l/s_T\right)^{q(1-\eta)}\int_{s_T/l}^{L/l}\left(L/l-f\right) f^{-q\eta}df\\
& = 4 C^q s_0^{q(\gamma-1)} L l^{q(2-\gamma)-1} \int_1^{s_T/l} f^{-q\gamma}df
\ -\  4 C^q s_0^{q(\gamma-1)} l^{q(2-\gamma)} \int_1^{s_T/l} f^{1-q\gamma} df\\
&\ +4 C^q s_0^{q(\gamma-1)} s_T^{q(\eta-\gamma)} L l^{q(2-\eta)-1} \int_{s_T/l}^{L/l} f^{-q\eta}df
\ -\  4C^q s_0^{q(\gamma-1)} s_T^{q(\eta-\gamma)} l^{q(2-\eta)}\int_{s_T/l}^{L/l} f^{1-q\eta}df\\
&=\begin{cases}
4 C^q s_0^{1-q} L l^{2(q-1)} \ln(s_T/l), &q\gamma = 1,\\
\frac{4 C^q s_0^{q(\gamma-1)}}{1-q\gamma}  L l^{q(2-\gamma)-1}\left[ \left(s_T/l\right)^{1-q\gamma} - 1 \right] , &q\gamma \neq 1,
\end{cases}\\ 
&\quad -
\begin{cases}
4 C^q s_0^{2-q} l^{2(q-1)} \ln (s_T/l), &q\gamma = 2,\\
\frac{4 C^q s_0^{q(\gamma-1)}}{2-q\gamma} l^{q(2-\gamma)} \left[\left( s_T/l \right)^{2-q\gamma} - 1 \right], &q\gamma \neq 2,
\end{cases} \\
&\ +
\begin{cases}
4 C^q s_0^{q(\gamma-1)} s_T^{1-q\gamma} L l^{2(q-1)} \ln(L/s_T), &q\eta = 1,\\
\frac{4 C^q s_0^{q(\gamma-1)} s_T^{1-q\gamma}}{1-q\eta} L l^{2(q-1)}\left[\left(L/s_T\right)^{1-q\eta} - 1 \right], &q\eta \neq 1,
\end{cases}\\
&\quad -
\begin{cases}
4 C^q s_0^{q(\gamma-1)} s_T^{2-q\gamma} l^{2(q-1)} \ln(L/s_T), &q\eta = 2,\\
\frac{4 C^q s_0^{q(\gamma-1)} s_T^{2-q\gamma}}{2-q\eta} l^{2(q-1)} \left[ \left(L/s_T \right)^{2-q\eta} - 1\right], &q\eta \neq 2,
\end{cases}\\
& \approx
\begin{cases}
\frac{4 C^q s_0^{q(\gamma-1)}}{q\gamma-1} L l^{q(2-\gamma)-1} \sim l^{q(2-\gamma)-1}, &q\gamma > 1,\\
4 C^q s_0^{1-q} L l^{2(q-1)} \ln(s_T/l) \sim l^{2(q-1)}, &q\gamma = 1,\\
\frac{4 C^q s_0^{q(\gamma-1)} s_T^{1-q\gamma} }{1-q\gamma} L l^{2(q-1)} \sim l^{2(q-1)}, &q\gamma < 1.
\end{cases}
\end{split}
\end{equation}
Interestingly, the dominant term in $Z_\text{off}(q,l)$ in Eq.~\ref{eq:double_detail_Zoff} depends on $\gamma$ rather than $\eta$.

\bibliographystyle{apsrev4-2}
%

\newpage
\begin{figure}
\centering
\includegraphics[width=\textwidth]{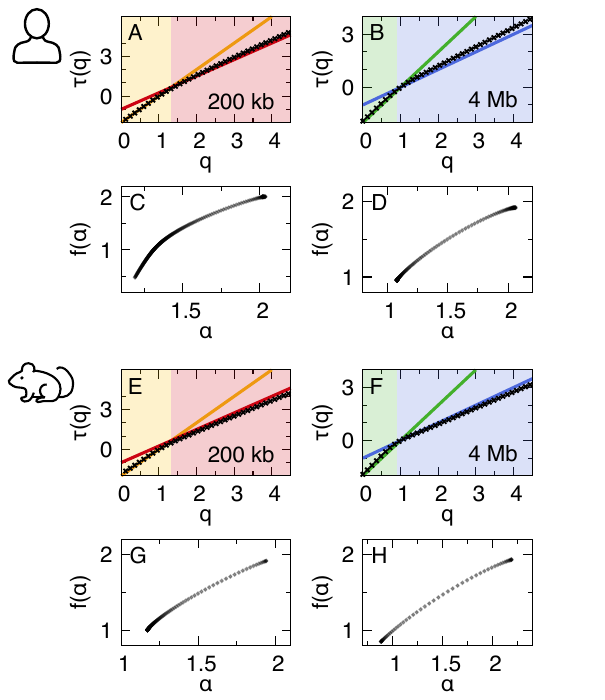}
\caption{
(A, B, E, and F) Mass exponent $\tau(q)$ and (C, D, G, and H) singularity spectrum $f(\alpha)$ for (A--D) human and (E--H) mouse Chr~1 Hi-C maps.
Singularity spectrum $f(\alpha)$ varies smoothly with $\alpha$, which implies that the system exhibits the multifractal behavior.
}\label{fig:MFS_Human_Mouse}
\end{figure}


\newpage
\begin{figure}
\centering
\includegraphics[width=\textwidth]{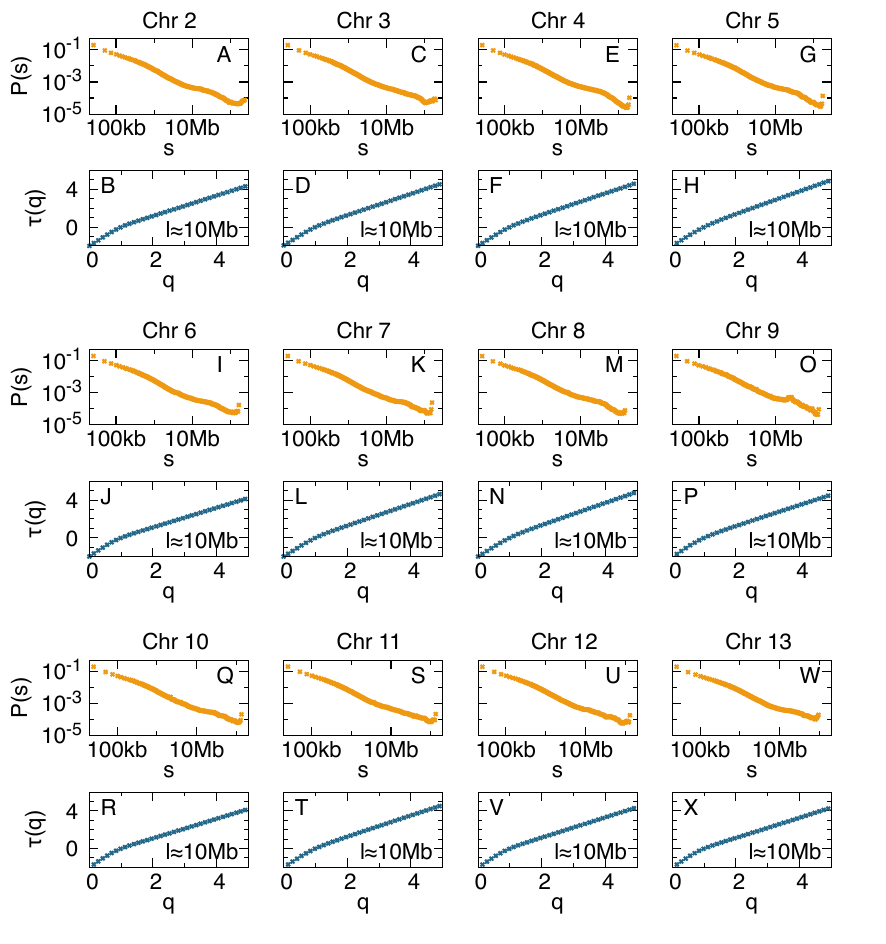}
\caption{
Contact probability $P(s)$ and multifractal analysis results for small and large $l$ for human Chr~2--Chr~13.
}\label{fig:Human_1}
\end{figure}


\newpage
\begin{figure}
\centering
\includegraphics[width=\textwidth]{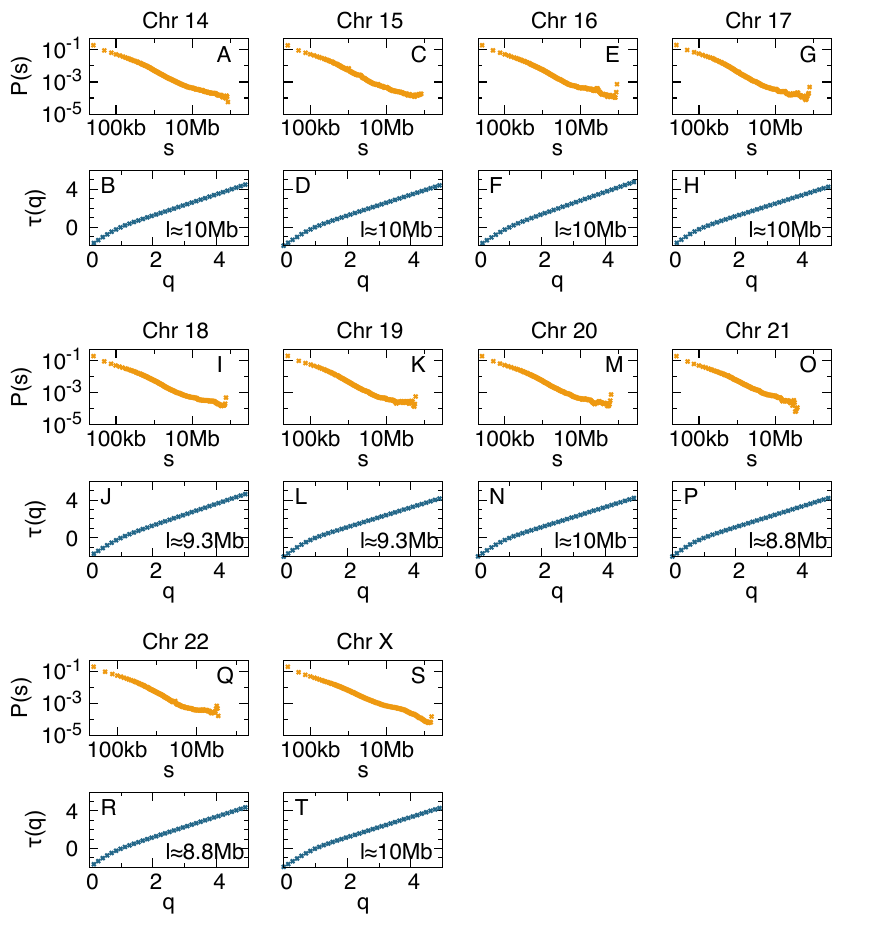}
\caption{
Contact probability $P(s)$ and multifractal analysis results for small and large $l$ for human Chr~14--Chr~X.
}\label{fig:Human_2}
\end{figure}


\newpage
\begin{figure}
\centering
\includegraphics[width=\textwidth]{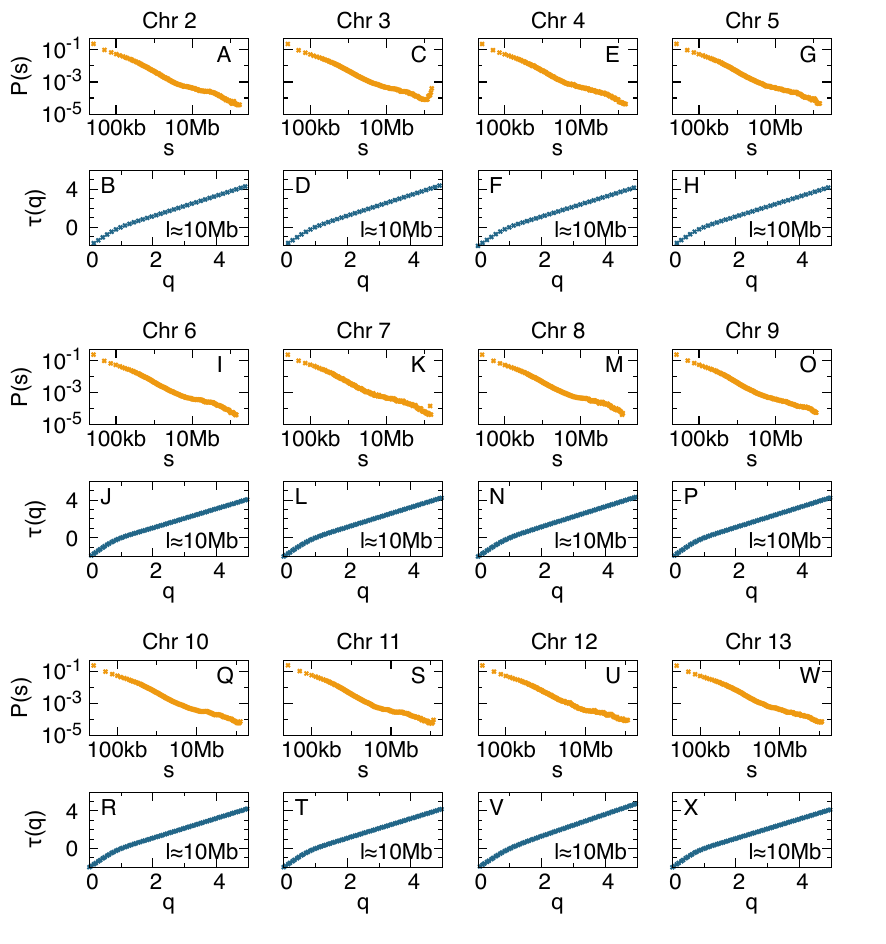}
\caption{
Contact probability $P(s)$ and multifractal analysis results for small and large $l$ for mouse Chr~2--Chr~13.
}\label{fig:Mouse_1}
\end{figure}


\newpage
\begin{figure}
\centering
\includegraphics[width=\textwidth]{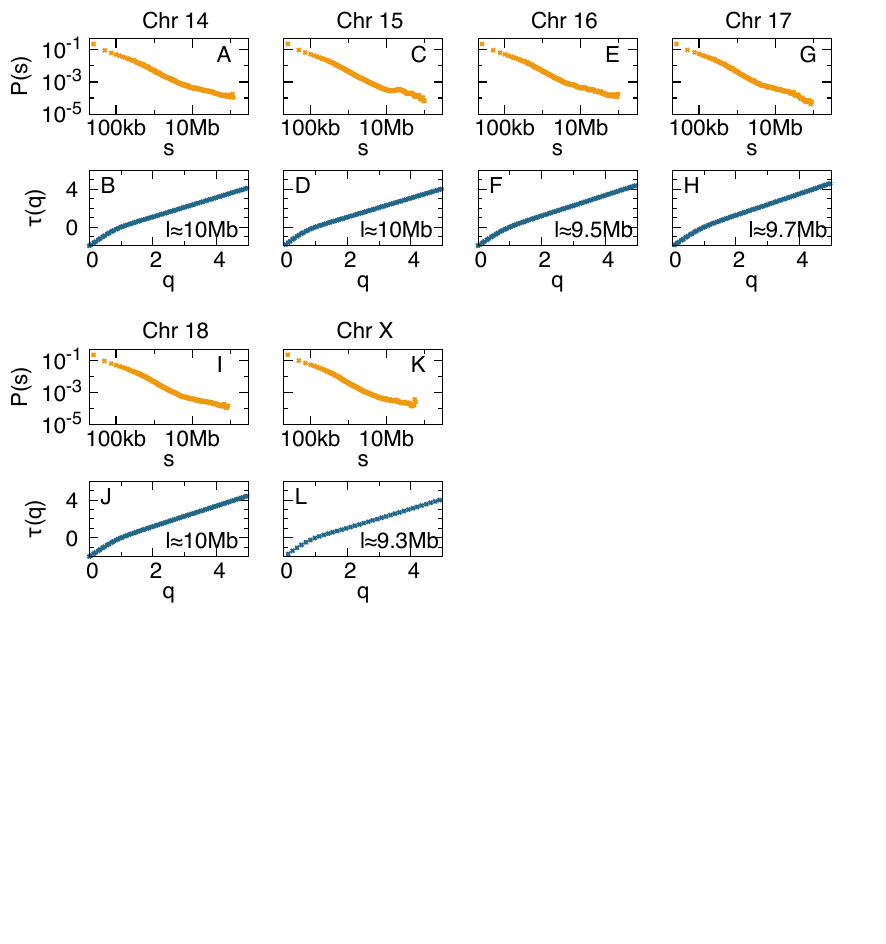}
\caption{
Contact probability $P(s)$ and multifractal analysis results for small and large $l$ for mouse Chr~14--Chr~X.
}\label{fig:Mouse_2}
\end{figure}


\newpage
\begin{figure}
\centering
\includegraphics[width=\textwidth]{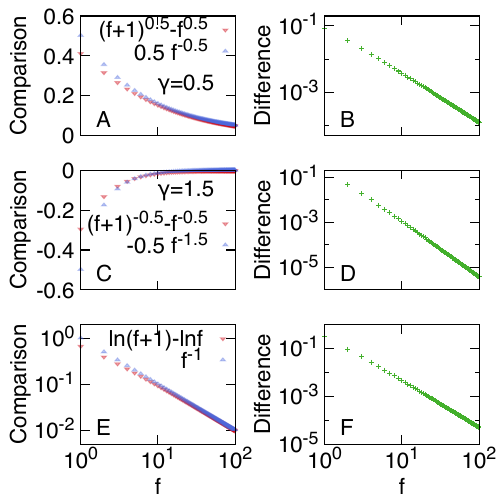}
\caption{
Approximations validation.
(A--D) We validate $(f+1)^{1-\gamma}-f^{1-\gamma}\approx (1-\gamma)f^{-\gamma}$ for (A and B) $\gamma=0.5$ and for (C and D) $\gamma=1.5$.
(E and F) Validation for $\ln(f+1) - \ln(f) \approx f^{-1}$.
(B, D, and F) As $f$ increase, the difference between original and approximated expressions decreases.
}\label{fig:Approx}
\end{figure}


\newpage
\begin{figure}
\centering
\includegraphics[width=\textwidth]{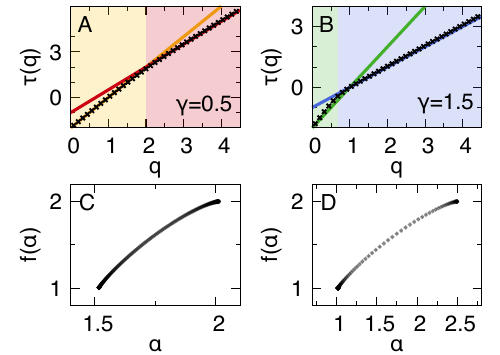}
\caption{
(A and B) Mass exponent $\tau(q)$ and (C and D) singularity spectrum $f(\alpha)$ for a single-exponent model.
The contact exponent $\gamma$ is (A and C) $\gamma=0.5$ and (B and D) $\gamma=1.5$.
}\label{fig:MFS_Single}
\end{figure}


\newpage
\begin{figure}
\centering
\includegraphics[width=\textwidth]{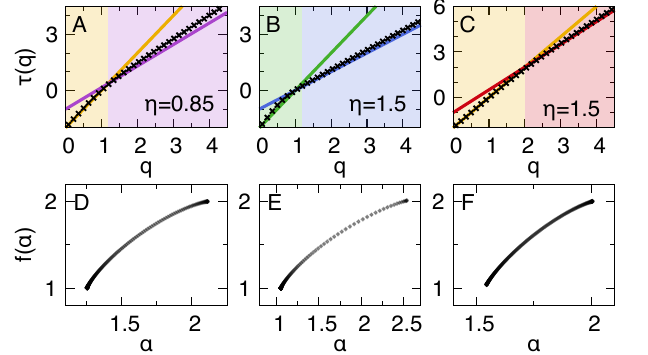}
\caption{
(A, B and C) Mass exponent $\tau(q)$ and (D, E and F) singularity spectrum $f(\alpha)$ for a double-exponent model.
The contact exponent $\eta$ at large-$s$ regime is (A and D) $\eta=0.85$ and (B, C, E, and F) $\eta=1.5$.
The other contact exponent $\gamma$ is set to $\gamma=0.5$.
}\label{fig:MFS_Double}
\end{figure}


\newpage
\begin{figure}
\centering
\includegraphics[width=\textwidth]{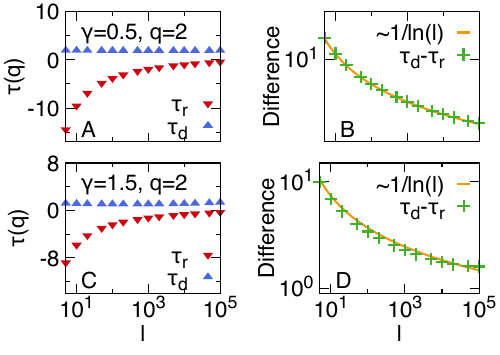}
\caption{
Comparison between the mass exponent $\tau(q)$ measured by ratio $\tau_r(q) = \ln Z(q,l) / \ln l$ and by log derivative $\tau_d(q) =  d\ln Z(q,l) / d\ln l$ for (A and B) $\gamma=0.5$ and (C and D) $\gamma=1.5$.
(A and C) Log-derivative result $\tau_d(q)$ shows more stable behavior than ratio result $\tau_r(q)$ for $l$.
(B and D) The difference $\tau_d(q) - \tau_r(q)$ scales as $1/\ln l$ as we expected.
We set $q=2$.
}\label{fig:DiffTau}
\end{figure}
